\documentclass[aps,prb,groupedaddress,superscriptaddress,twocolumn]{revtex4-2}
\usepackage{amsmath,graphicx}
\usepackage{float} 
\usepackage{subfigure} 
\usepackage{color}
\usepackage{braket}
\usepackage{mathrsfs}
\usepackage{appendix}

\begin{document}
\title{Anomalous photon-assisted tunneling in periodically driven Majorana nanowires and BCS Charge Measurement}
\author{Yuchen Zhuang}
\affiliation{International Center for Quantum Materials, School of Physics, Peking University, Beijing 100871, China}
\affiliation{CAS Center for Excellence in Topological Quantum Computation, University of Chinese Academy of Sciences, Beijing 100190, China}

\author{Qing-Feng Sun}
\email[]{sunqf@pku.edu.cn}
\affiliation{International Center for Quantum Materials, School of Physics, Peking University, Beijing 100871, China}
\affiliation{CAS Center for Excellence in Topological Quantum Computation, University of Chinese Academy of Sciences, Beijing 100190, China}
\affiliation{Beijing Academy of Quantum Information Sciences, West Bld.\#3, No.10 Xibeiwang East Rd., Haidian District, Beijing 100193, China}

\date{\today}

\begin{abstract}
The photon-assisted tunneling of Majorana bound states
in a Majorana nanowire driven by the periodic field is studied
both theoretically and numerically. We find that
Majorana bound states exhibit an anomalous photon-assisted
tunneling signal which is different from an ordinary fermionic state : the height of photonic sidebands is related to the degree of Majorana nonlocality.
Moreover, we show that the Bardeen-Cooper-Schrieffer (BCS) charge and spin components of subgap states
can be well revealed by the local conductance.
Our work illuminates the effect of driving field on Majorana bound states and provides a systematic scheme to directly measure BCS information of overlapping Majorana bound states.
\end{abstract}

\maketitle	
\section{\label{sec1} Introduction}
A Majorana bound state (MBS) is a chargeless quasiparticle
whose antiparticle is equal to itself  \cite{Wilczek, Kitaev, Alicea, Elliott}.
It has attracted a widespread attention for a long time because of its potential applications in fault-tolerant quantum computations \cite{Kitaev2, Nayak, Aasen}.
One way to prepare MBSs is based on an one-dimensional (1D)
hybrid superconducting-semiconducting nanowire in the presence of spin-orbit coupling and Zeeman splitting (often dubbed as Majorana nanowire) \cite{Lutchyn,Oreg,Mourik}. MBSs can lead to a series of novel transport phenomena, such as a quantized zero bias conductance peak (ZBCP) with value of $2e^{2}/h$ through the resonant Andreev reflection \cite{Law}, fractional Josephson effect \cite{Kitaev, Fu} and spin selective equal-spin Andreev reflection \cite{He}.

Although an isolated zero-energy MBS has no charge and spin \cite{Pawlak}, when two MBSs are coupled (i.e. Majorana nanowire's length $L$ is not longer enough than Majorana decay length $\xi_M$ ), a nonzero energy state with non-quantized charge (BCS charge) and spin will be generated \cite{Lin,Ben,Fernando}. This finite BCS charge is related to the nonlocality of MBSs \cite{Prada1,Deng,Penaranda}, which is of particular significance to implement a topological qubit \cite{Deng}. Recently, both theory and experiment have demonstrated that a three-terminal set-up can be used to extract the local BCS charge of bound states close to the ends of the wire \cite{Danon, Menard}. However, a systematic scheme to measure this BCS charge (e.g. measure its spin and spatial information) is still lacking.

Periodically driven systems have been concerned about for a long time \cite{Tucker,Platero}. One of the key features is that electrons tunneling through these systems can absorb or
emit multiple photons, causing photonic sideband peaks at harmonics in the conductance-voltage curves, which is known as photon-assisted tunneling (PAT). The study of this phenomenon can trace back to the early pioneering work by Tien and Gordon in superconductor-insulator-superconductor films \cite{Tien}. They raise a simple relation that n-th PAT peaks is modulated by a square of Bessel function $J^{2}_{n}\left(eV_{f}/ \hbar \omega\right)$ \cite{Kot}. Here $e$ is charge of the electron, $\hbar \omega$ is the photon's energy and $V_{f}$ is the driving amplitude. Until now, periodically-driving physics has been explored in a widespread range including quantum dots \cite{Kouwenhoven,Sun3}, semiconductor superlattice \cite{Keay},
as well as two-dimensional electron gas \cite{Mani,Zudov,Shi}.

The interplay between PAT and MBSs in the topological superconductor also raises recent common interest. One recent work has studied the PAT from ac-driven normal electrodes into the Majorana nanowire with MBSs and find resulting nonzero sideband peaks \cite{Tang}. Another recent work has analyzed the photon-assisted resonant Andreev reflection from ac-driven superconducting tips or normal electrodes into subgap states like Yu-Shiba-Rusinov (YSR) states and MBSs, and argue that it could provide a high-accuracy method to measure small but nonzero energies of subgap states \cite{Gonzalez}. But essentially, these previous studies focus on a case that complete fermions are driven to absorb and emit photons tunneling into the subgap states in superconducting system. But, how MBS, as a `half fermion', responses under the periodically driving field is still an open question. Moreover, since PAT signal is directly connected with the charge in the tunneling process, we can naturally wonder that whether PAT could give us the BCS charge information of bound states.

\begin{figure}[ht]
	\includegraphics[width=1\columnwidth]{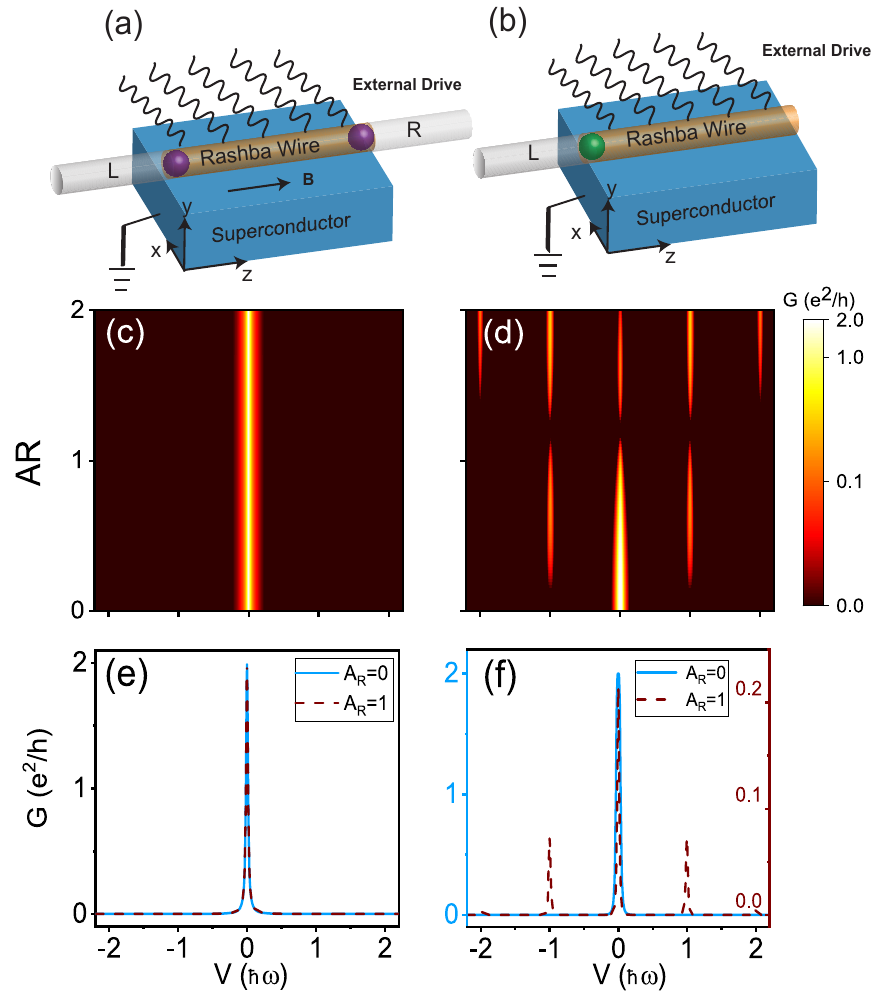}
	\centering 
	\caption{(a, b)
	Schematic representation for MBSs (a) and single impurity (b)
	in periodically-driven nanowire which is in the proximity
	to the grounded superconductor.
	Two MBSs (purple dots) locate at the two ends of the 1D Majorana nanowire
	which is connected to two normal leads (a).
	A trivial impurity (green dot) lies between the normal left lead and the semiconductor nanowire (b).
    (c-f) Time-averaged local conductance $G$ as
    functions of bias $V$ and relative driving field strength $A_{R}$ for the MBSs (c, e) and for the trivial impurity (d, f). Here nanowire length $N=700$, static magnetic field $V_{0}=400\mu eV$, chemical potential $\mu=0$, the photon energy $\hbar \omega=20\mu eV$, corresponding to frequency $f\approx 5GHz$. Other parameters have been given in the text. }
	\label{FIG1}
\end{figure}

The goal of this paper is to study the PAT signals of MBSs in the periodically-driven
Majorana nanowire connected by two normal leads,
in proximity to an equilibrium superconductor, as shown in Fig.~\ref{FIG1}(a). Both numerically and analytically, our results illuminate that MBSs show an anomalous PAT signal different from an ordinary fermionic state in Fig.~\ref{FIG1}(b): photonic sideband peaks disappear when MBSs are well-separated but reappear once MBSs are coupled. This phenomenon is attributed to the charge-neutral and spinless properties of Majorana fermions.
We further prove that the height of PAT peaks is relevant to the BCS charge of subgap states. Based on this, we propose a method to directly extract both BCS charge and spin components of overlapping MBSs just via the local conductance.

The rest of this paper is organized as follows.
In Sec II, we give our model and present
a low-energy effective model to describe the MBSs under the periodically-driving field.
In Sec III, we use nonequilibrium Green's function method to calculate current and conductance. We also conduct some numerical calculations for MBSs and  trivial Andreev bound states (ABSs) induced by hybridized impurity respectively, and further compare their results.
In Sec IV, we analyze the time-averaged conductance of overlapping Majoranas
and point out one way to extract their BCS charge.
In Sec V, we consider a case where microwave field and harmonic magnetic field coexist and discuss how to extract the BCS spin information along any direction.
We put the summary and some discussions in Sec VI. Some detailed calculations and supplementary figures are all concluded in Appendices A-C.

\section{\label{sec2} The model and Hamiltonian}
\subsection*{A. Model for periodically driven Majorana nanowire and single impurity}
Specifically, to describe the 1D Majorana nanowires
as shown in Fig.~\ref{FIG1}(a),
the Bogoliubov-de Gennes (BdG)
Hamiltonian in the Nambu basis $\{ \psi_{\uparrow}(z),\psi_{\downarrow}(z),\psi^{\dagger}_{\uparrow}(z),\psi^{\dagger}_{\downarrow}(z)\} $ can be written as \cite{Lutchyn,Oreg}

\begin{equation}
H_{nw}^{BdG}=(\frac{p_{z}^{2}}{2m^{*}}-\mu+V_{0}\sigma^{z}+\alpha_{R}p_{z}\sigma^{y})\tau^{z}-\widetilde{\Delta} \sigma^{y}\tau^{y},
\label{Eq1}
\end{equation}
where Pauli matrices $\sigma$ and $\tau$ act on the spin and particle-hole space.
$m^{*}$ is the effective electronic mass.
$\alpha_{R} $ and $V_{0}=g\mu_{B}B/2$ are spin-orbit coupling strength
and the static Zeeman field with land$\acute{e}$ $g$-factor
and magnetic field $B$,
$\mu$ and $\tilde{\Delta}$ represent the chemical potential
and proximity-induced pairing potential. To facilitate numerical calculations,
we discretize this Hamiltonian into tight-binding form and also include the periodically-driving field and coupling between left and right leads:

\begin{equation}
H_{MBS}\left(t\right)=H_{nw}+H_{fm}\left(t\right)+H_{T1}+H_{L}+H_{R},
\label{Eq2}
\end{equation}
where
\begin{equation}
	\begin{split}
	H_{nw}&=\sum_{i=1,s}^{N}( 2t_{N}-\mu)c^{\dagger}_{is}c_{is}
	-\sum_{i=1,s}^{N-1} t_{N}(c^{\dagger}_{is}c_{i+1s}+h.c.)\\
	&-\sum_{i=1,s,s'}^{N-1} \left[\frac{\alpha_R}{2a}c^{\dagger}_{is}
	(i\sigma^{y})_{ss'}c_{i+1s'}+h.c.\right]\\
	&+\sum_{i=1,s,s'}^{N} V_{0}c^{\dagger}_{is}(\sigma^{z})_{ss'}c_{is'}
	+\sum_{i=1}^{N}(\widetilde{\Delta} c^{\dagger}_{i\uparrow}c^{\dagger}_{i\downarrow}+h.c.)
	\end{split}
	\label{Eq3}
	\end{equation}
and
\begin{equation}
	\begin{split}
		H_{fm}\left(t\right)&=\sum_{i,s}A \cos (\omega t)c_{is}^{\dagger}c_{is},\\
		H_{\alpha=L/R}&=\sum_{k\alpha,s}\varepsilon_{k\alpha}a_{k\alpha s}^{\dagger}a_{k\alpha s},\\
		H_{T1}&=\sum_{kL,s}t_{L}c_{1s}^{\dagger}a_{kLs}
+\sum_{kRs}t_{R}c_{Ns}^{\dagger}a_{kRs}+h.c.
	\end{split}
	\label{Eq4}
	\end{equation}
Here $i$ and $s$ ($s=\uparrow,\downarrow$ or $\pm 1$)
label the lattice coordinate and the spin index,
$N$ and $a$ is the number of lattice points
and the lattice constant of Majorana nanowire,
$t_{N}$ is the nearest hopping energy.
$H_{fm}$ describes the microwave driving field with the driving amplitude $A$ . Besides microwave field, harmonic magnetic field along z direction is also investigated in our following calculations with $H_{fz}\left(t\right)=\sum_{i,ss'}Acos\left(\omega t\right)\left(\sigma^{z}\right) _{ss'}c_{is}^{\dagger}c_{is'}$.
$ H_{\alpha=L/R}$ describes left/right normal leads ($t_{L,R}$ can be set as real).
$H_{T1}$
is the tunneling Hamiltonian between central Majorana nanowires and left/right leads.
For simplicity, we have made some assumptions:
(i) The external harmonic field is only applied on the 1D Majorana nanowire and does not affect other parts \cite{Liu2,Yang}.
(ii) The frequency of the external field is much lower than
$\tilde{\Delta}$ and the adiabatic approximation holds,
thus the harmonic field only changes the single electron energy \cite{Wingreen,Sun1,Sun2}. In fact, even if frequency reaches to $10$GHz, the energy $\hbar\omega$ is about 40$\mu$eV.
This is usually much smaller than $\tilde{\Delta}$ which is around 200 $\mu$eV.

To compare the MBS with an ordinary fermionic state, we consider another simple case where an impurity is  hybridized with the superconductor [Fig.~\ref{FIG1}(b)]. This kind of ABSs can be described by the impurity Hamiltonian \cite{Baran}
\begin{equation}
H_{ABS}=H_{d}+H_{sc} +H_{T2}+H_{L}+H_{fm}\left(t\right) ,
\label{Eq5}
\end{equation}
where
\begin{equation}
	\begin{split}
	H_{d}&=\sum_{s}E_{0}d^{\dagger}_{s}d_{s},\\
	H_{sc}&=\sum_{k,s}\varepsilon_{sc,k}b^{\dagger}_{ks}b_{ks}
+\sum_{k}\tilde{\Delta} b^{\dagger}_{k \uparrow}b^{\dagger}_{-k \downarrow}+H.c.,\\
H_{fm}\left(t\right)&=\sum_{s}Acos\left(wt\right)d^{\dagger}_{s}d_{s},\\
    H_{T2}&=\sum_{k s}t_{L}d_{s}^{\dagger}a_{kL s}+t_{sc}d_{s}^{\dagger}b_{ks}+H.c.
	\end{split}
	\label{Eq6}
	\end{equation}
Here $H_{d}$ is the impurity Hamiltonian. $H_{sc}$ is the grounded s-wave superconductor Hamiltonian. $H_{fm}$ and $H_{T2}$ is for the microwave driving field and tunneling part respectively ($H_{fz}$ is also included in the following).
The Hamiltonian $H_{L}$ for the normal left lead are the same as the former case.
Based on intrinsic superconducting limit and wide-band limit \cite{Jauho,Yang},
the superconductor reservoir can be integrated out to introduce
a self-energy term $\mathbf{\Sigma} _{sc}^{r}=-\frac{\widetilde{\Gamma}^{sc}}{2}d_{\uparrow}^{\dagger }d_{\downarrow}^{\dagger}+h.c.$ with $\widetilde{\Gamma}^{sc}=2\pi\rho_{s}\left | t_{sc} \right |^{2}$ ($\rho_{s}$ is the superconducting density of states) \cite{Yang, Baran}.
Similarly, the couplings between Majorana nanowire
and the normal leads can also introduce a self-energy term $\Sigma^{r}_{L,ij,lr}=-i\frac{\widetilde{\Gamma}^{L}}{2}\delta_{ij}\delta_{j1}\delta_{lr}$ and $\Sigma^{r}_{R,ij,lr}=-i\frac{\widetilde{\Gamma}^{R}}{2}\delta_{ij}\delta_{jN}\delta_{lr}$ with $\widetilde{\Gamma}^{\alpha=L,R}=2\pi\rho_{\alpha}\left | t_{\alpha} \right |^{2}$ ($\rho_{\alpha}$ is the density of states in the normal lead $\alpha$ ). Here the first two indices denote sites and the other two indices denote Nambu spinor.

\subsection*{B. Projected Hamiltonian and BCS charge}
In order to analytically investigate
how the MBSs are affected by microwave driving field, by Bogoliubov transformation $\psi^{\dagger}_{n}
=\sum_{i,s} u_{is}^{(n)}c^{\dagger}_{is}+v_{is}^{(n)}c_{is}$,
Majorana nanowire Hamiltonian $H_{nm}$ in Eq.(\ref{Eq3}) can be diagonalized
as
\begin{equation}
 H_{nw}=\sum_{n} \varepsilon_{n}\psi^{\dagger}_{n}\psi_{n}.
\label{Eq7}
\end{equation}
$ u_{is}^{(n)}, v_{is}^{(n)} $ are the particle and hole components
of the $\varepsilon_{n}$ eigenstate at the site $i$ with spin $s$.
In view that the energy of concerned subgap states and driving field frequency $\omega$
are both much lower than the induced topological gap,
the mix between subgap states and quasiparticle continuum could be ignored.
Hence, we can just focus on the projected Hamiltonians
of Majorana nanowire and harmonic field, $H^{p} =H_{nm}^{p} +H_{fm/z}^{p}$,
in the space formed by the lowest energy subgap state
$(\psi^{\dagger}_{0},\psi_{0})$ with energy $\epsilon_{0}$:
\begin{equation}
\begin{split}
&P=\begin{pmatrix}
		u_{1\uparrow}^{(0)*} & u_{1\downarrow}^{(0)*} & v_{1\uparrow}^{(0)*}& v_{1\downarrow}^{(0)*} \cdots & u_{N\uparrow}^{(0)*} & u_{N\downarrow}^{(0)*}& v_{N\uparrow}^{(0)*} & v_{N\downarrow}^{(0)*}\\
		v_{1\uparrow}^{(0)} & v_{1\downarrow}^{(0)} & u_{1\uparrow}^{(0)} & u_{1\downarrow}^{(0)}  \cdots & v_{N\uparrow}^{(0)} & v_{N\downarrow}^{(0)} & u_{N\uparrow}^{(0)} & u_{N\downarrow}^{(0)}
		\end{pmatrix},\\
&H_{nw}^{p}=PH_{nw}P^{\dagger}=
\begin{pmatrix}
\varepsilon_{0} & 0\\
0 & -\varepsilon_{0}
\end{pmatrix},\\
&H_{fm}^{p}=PH_{fm}P^{\dagger}=
\begin{pmatrix}
QAcos\left(wt\right) & 0\\
0 & -QAcos\left(wt\right)
\end{pmatrix},\\
&H_{fz}^{p}
=PH_{fz}P^{\dagger}=\begin{pmatrix}
	\zeta_{z} A\cos\left( \omega t\right)  & 0 \\
	0 & -\zeta_{z} A\cos\left ( \omega t\right)
	\end{pmatrix}.
\label{Eq8}
\end{split}
\end{equation}
$P$ is the projection operator on the lowest energy states. $Q$ and $\zeta_{z}$ is the BCS charge and BCS spin polarization along z direction for the quasi-particle state $\psi_{0}^{\dagger}$:
\begin{equation}
	\begin{split}
Q&=\sum_{i,s} \left(\left|u_{is}^{(0)} \right |^{2}
-\left | v_{is}^{(0)} \right |^{2}\right),\\
\zeta_{z} &=\sum_{i,s} s\left ( \left | u_{is}^{(0)} \right |^{2}
-\left | v_{is}^{(0)} \right |^{2}\right ).
	\end{split}
\label{Eq9}
\end{equation}
In the continuum space, $Q$ or $\zeta$ could be generalized as $Q=\sum_{s} \int \left(\left|u_{s}^{(0)}(z) \right |^{2}
-\left | v_{s}^{(0)}(z) \right |^{2}\right) \,dz$ or $\zeta_{z}=\sum_{s} s\int \left(\left|u_{s}^{(0)}(z) \right |^{2}
-\left | v_{s}^{(0)}(z) \right |^{2}\right) \,dz$.
Equation (\ref{Eq8}) reveals a fact:
the external driving field on the lowest energy state of $H_{nm}$
in our case should be renormalized by a factor, corresponding to the BCS charge Q (or BCS spin polarization $\zeta_{z}$).
It's natural in view of quasi-particles interacting
with external fields with charge $Qe$ rather than $e$ as electrons.
For the further analysis, we also define its magnitude as
\begin{equation}
	Q_{BCS}=\left| \sum_{i,s} \left(\left|u_{is}^{(0)} \right |^{2}
-\left | v_{is}^{(0)} \right |^{2}\right)\right|
\label{Eq10}
\end{equation}
and BCS spin component magnitude as
\begin{equation}
Q_{BCS,s}=\left|\sum_{i}\left(\left | u_{is}^{(0)} \right |^{2}
-\left | v_{is}^{(0)} \right |^{2}\right)\right|.
\label{Eq11}
\end{equation}
$Q_{BCS}$ here is parallel to the definition
of $\delta N$ introduced in Refs. \cite{Ben, Penaranda}. Its value ranges from 0 to 1.

It is easy to find that $Q=0$ and $\zeta_{z}=0$ as long as $\psi^{\dagger}=\psi$, which echoes Majoranas' charge-neutral and spinless properties. Thus, isolated MBS itself should not be affected by the driving field. For a sufficiently long 1D Majorana nanowire in topological region,
the lowest energy states are nonlocal states formed by two well-separated MBSs at ends of wire, with zero BCS charge.
As a result, PAT signals must disappear in this case.
In contrast, the fermionic trivial ABSs are usually charged, even though their energy crosses zero. PAT signals for them should be observed.

The deeper meaning of the BCS charge is related
to the Majorana nonlocality \cite{Prada1,Penaranda,Deng}. This can be proved as follows. For any zero-energy ABS, regardless of its topological origin, can always be decomposed into two MBSs:
\begin{equation}
	\gamma_{1}=\psi_{0}^{\dagger}+\psi_{0}, \quad   \gamma_{2}=i( \psi_{0}^{\dagger}-\psi_{0}).
	\label{Eq12}
\end{equation}
The spatial wavefunction of MBS can be written as  $\gamma_{i}
=\sum_{j,s} \Phi_{js}^{(\gamma_{i})}c^{\dagger}_{js}+(\Phi_{js}^{(\gamma_{i})})^{*}c_{js}$, with
\begin{equation}
\begin{split}	
	\Phi_{js}^{\gamma_{1}}&=u_{js}^{(0)}+(v_{js}^{(0)})^{*}\\
	\Phi_{js}^{\gamma_{2}}&=i(u_{js}^{(0)}-(v_{js}^{(0)})^{*}).
\end{split}
\label{Eq13}
\end{equation}
Without lose of generality, $u_{is}^{(0)}$ and $v_{is}^{(0)}$ can be set as real and BCS charge magnitude in Eq. (\ref{Eq10}) could be regulated as overlap between MBSs' wavefunctions :
\begin{equation}
   Q_{BCS}=\left|\sum_{js}\Phi_{js}^{\gamma_{1}}*\Phi_{js}^{\gamma_{2}}\right|.
   \label{Eq14}
\end{equation}
Trivial ABSs are composed of two highly-overlapping or partially separated MBSs with finite BCS charge. Usually, when MBSs are spatially-separated (e.g. topological MBSs located at two ends of the nanowire) or spin-separated (e.g. quasi-MBSs originating from two spin channels \cite{Kells, Vuik}), BCS charge in Eq.~(\ref{Eq14}) can approach zero within a large parameter range.
This kind of well-separated MBSs have potential value in (parametric) non-Abelian braiding \cite{Vuik, Prada2}. We reveal that BCS charge information of overlapping MBSs can naturally appear in Eq. (\ref{Eq8}). This is one of our main findings.

\section{\label{sec3} Currents and Conductance under Periodically-driving field}
In order to demonstrate our argument in the section II. In this section, we try to derive the current and conductance in detail for periodically-driven nanowires and impurity by using nonequilibrium Green's function method \cite{Jauho, Wingreen, Sun1,Sun2}. Along with Floquet-Landauer formalism \cite{ Carlos,Rudner,Kohler}, numerical results are also performed.
\subsection*{A. Derivation of current and conductance}
Assuming the bias voltage is only applied on the left lead with $V_{L}=V$
and the others are connected to the ground.
The current flowing from the normal left lead into
the periodically-driven Majorana nanowire
via the time derivative of electron number operator ($N_{L}=\sum_{k,s}a^{\dagger}_{kL,s}a_{kL,s}$) is : \cite{Wingreen, Sun2}
\begin{equation}
\begin{split}
	I_L\left ( t \right )&= -e \langle \dot{N}_{L} \rangle
	=2eRe\left \{ \sum_{k,l(l=1,2)}t_{L}G_{1kL,ll}^{<}\left ( t,t \right ) \right \}.
	\end{split}
	\label{15}
	\end{equation}
	Here the Green's functions $\mathbf{G}^{<}, \mathbf{G}^{r} $ are defined in 4-component Nambu basis as
	\begin{equation}
	\begin{split}
	\textbf{G}_{ij}^{<}&(t,t')
	\equiv i \langle \begin{pmatrix}
	c_{j\uparrow}^{\dagger}(t^{'})\\c_{j\downarrow}^{\dagger}(t^{'})
	\\c_{j\uparrow}(t^{'})
	\\c_{j\downarrow} (t^{'})
	\end{pmatrix}
	\begin{pmatrix}
	c_{i\uparrow}(t),&c_{i\downarrow}(t),&c_{i\uparrow}^{\dagger}(t),&c_{i\downarrow}^{\dagger}(t)
	\end{pmatrix} \rangle,\\
	\textbf{G}_{ij}^{r}&(t,t')
	\equiv -i\theta (t-t') \\
	&\times \langle \{ \begin{pmatrix}
	c_{i\uparrow}(t)\\c_{i\downarrow}(t)
	\\c_{i\uparrow}^{\dagger}(t)
	\\c_{i\downarrow}^{\dagger}(t)
	\end{pmatrix},
	\begin{pmatrix}
	c_{j\uparrow}^{\dagger}(t'),&c_{j\downarrow}^{\dagger}(t'),&c_{j\uparrow}(t'),&c_{j\downarrow}(t')
	\end{pmatrix}\} \rangle
	\end{split}
	\label{S5}
	\end{equation}
where $i,j$ represent the index $kL$, $kR$ in the normal leads,
and site $1, 2, 3,$...$N$ in Majorana nanowire.
Using Dyson equation \cite{Wingreen,Sun2} and wide-band limit \cite{Jauho}, the current can be formulated as (in units of $\hbar=1$):
\begin{equation}
	\begin{split}
	&I_L ( t )=\sum_{l=1}^{2}\left(-2e\int_{-\infty}^{t}dt_{1}\int\frac{d\varepsilon}{2\pi}\widetilde{\Gamma} ^{L}f_{L}\left (\varepsilon\right)\right.\\
	&\left.Im \left \{e^{-i\varepsilon\left ( t_{1}-t\right )}G^{r}_{11,ll}\left (t,t_{1}\right)\right \}-e\widetilde{\Gamma} ^{L} Im\left\{G^{<}_{11,ll}\left(t,t\right)\right \}\right)
	\end{split}
	\label{Eq17}
	\end{equation}
with the first two subscripts of the Green's functions denoting sites and the second two subscripts denoting Nambu spinor. We introduce level-width function matrices $(\mathbf{\Gamma}^{L})_{ij,lr}
=2\pi\rho_{L}t_{L}^{2}\delta_{ij}\delta_{j1}\delta_{lr}
\equiv \widetilde{\Gamma}^{L} \delta_{ij}\delta_{j1}\delta_{lr}$
and
$(\mathbf{\Gamma}^{R})_{ij,lr}
=2\pi\rho_{R}t_{R}^{2}\delta_{ij}\delta_{jN}\delta_{lr}
\equiv \widetilde{\Gamma}^{R}\delta_{ij}\delta_{jN}\delta_{lr}$. To relate $\mathbf{G}^{<}$ and $\mathbf{G}^{r}$, use Keldysh equation \cite{Jauho}:
\begin{equation}
	\mathbf{G}^{<}(t,t')=\int dt_{1} \int dt_{2} \mathbf{G}^{r}(t,t_{1})\mathbf{\Sigma}^{<}(t_{1},t_{2})\mathbf{G}^{a}(t_{2},t').
	\label{Eq18}
\end{equation}
Lesser self-energy due to the coupling between the left/right lead
and the Majorana nanowire is
$\mathbf{\Sigma}^{<}(t_{1},t_{2})=i\sum_{\alpha=L,R}
\int \frac{d\varepsilon }{2\pi }e^{-i\varepsilon(t_{1}-t_{2})}\mathbf{f}_{\alpha}(\varepsilon )
\mathbf{\Gamma}^{\alpha}$ with
$\mathbf{f}_{\alpha}(\varepsilon)=diag( f_{\alpha e}(\varepsilon ), f_{\alpha e}(\varepsilon ), f_{\alpha h}(\varepsilon ), f_{\alpha h}(\varepsilon ))$.
$f_{\alpha e}\left( \varepsilon\right)=
\left(e^{\frac{\varepsilon-eV_{\alpha}}{k_{B} T}}+1\right)^{-1} $
and
$f_{\alpha h}\left( \varepsilon\right)=
\left(e^{\frac{\varepsilon +eV_{\alpha}}{k_{B} T}}+1\right)^{-1} $
are Fermi-Dirac distribution of electrons and holes. Substituting Eq. (\ref{Eq18}) into Eq. (\ref{Eq17}), we can get current:
\begin{widetext}
\begin{equation}
	\begin{split}
		I_L \left ( t \right )
		&=\sum_{l=1}^{2}
		\left\{-2e\int \frac{d\varepsilon}{2\pi}\widetilde{\Gamma}^{L}
		f_{Le}\left(\varepsilon\right) Im \left[ \mathscr{\mathscr{A}}_{11,ll}
		\left( \varepsilon,t \right)\right]
		 -e\left(\widetilde{\Gamma}^{L}\right)^{2} \int\frac{d\varepsilon}{2\pi}
		\left[ f_{Le} \left (\varepsilon\right) \sum_{r=1}^{2}
		\left| \mathscr{\mathscr{A}}_{11;lr}\left(\varepsilon,t \right)\right|^{2}+
		f_{Lh}\left(\varepsilon\right)\sum_{r=3}^{4}\left|\mathscr{A}_{11;lr}
		\left(\varepsilon,t \right)\right|^{2}\right] \right.\\
		& \left.-e\widetilde{\Gamma}^{L}\widetilde{\Gamma}^{R}
		\int\frac{d\varepsilon}{2\pi} \left[
		f_{Re}\left(\varepsilon\right)\sum_{r=1}^{2}\left| \mathscr{A}_{1N;lr}
		\left(\varepsilon,t \right)\right|^{2}+ f_{Rh}(\varepsilon)
		\sum_{r=3}^{4}\left|\mathscr{A}_{1N;lr}\left(\varepsilon,t\right)\right|^{2}
		\right] \right\}.
	\end{split}
	\label{Eq19}
\end{equation}	

Notation $\mathscr{A}$ is defined as
\begin{equation}
\mathbf{\mathscr{A}}\left(\varepsilon,t\right)=\int dt_{1}\mathbf{G}^{r}\left (t,t_{1}\right)e^{i\varepsilon\left ( t-t_{1}\right )}.
\label{Eq20}
\end{equation}
If the system is time-independent, $\mathbf{\mathscr{A}}$ actually refers to the Fourier transformation of retarded Green function $\mathbf{G}^{r}$. In case of the time-periodic potential,
we can refer to a Fourier transformation defined as \cite{Sun1, Li}
\begin{equation}
\mathbf{G}\left ( t,t_{1} \right )=\sum_{n}e^{in\omega t_{1}}\int \frac{d\varepsilon}{2\pi}e^{-i\varepsilon\left ( t-t_{1}\right )}\mathbf{G}_{n}\left (\varepsilon\right ).
\label{Eq21}
\end{equation}
We can also introduce the notation $\mathbf{G}_{mn}\left(\varepsilon \right)=\mathbf{G}_{n-m}\left(\varepsilon+m\omega \right)$ to relate different $\mathbf{G}_{mn}$ components. Then $\mathbf{G}_{mn}(\varepsilon)=\mathbf{G}_{0,n-m}\left(\varepsilon+m\omega \right)$.
Under this transformation, the current in Eq.~(\ref{Eq19}) can be reworded as

\begin{equation}
	\begin{split}
	I_L(t)
	&=\sum_{l=1}^{2}\left\{
	-2e\int \frac{d\varepsilon}{2\pi}\widetilde{\Gamma} ^{L}f_{Le}\left (\varepsilon\right)Im\left[ \sum_{n}e^{in\omega t}G^{r}_{11,ll;-n0}\left ( \varepsilon\right )\right] \right.\\
	& -e\left(\widetilde{\Gamma} ^{L}\right)^{2}
	\int\frac{d\varepsilon}{2\pi}\left[ f_{Le}\left (\varepsilon\right)\sum_{r=1}^{2}\left| \sum_{n}e^{in\omega t}G^{r}_{11,lr;-n0}\left ( \varepsilon\right)\right|^{2}
	+f_{Lh}\left (\varepsilon\right)
	\sum_{r=3}^{4}\left|\sum_{n}e^{in\omega t}G^{r}_{11,lr;-n0}
	\left ( \varepsilon\right )\right|^{2}\right]\\
	& \left.-e\widetilde{\Gamma} ^{L}\widetilde{\Gamma} ^{R} \int\frac{d\varepsilon}{2\pi}\left[ f_{Re}\left (\varepsilon\right)\sum_{r=1}^{2}\left| \sum_{n}e^{in\omega t}G^{r}_{1N,lr;-n0}\left ( \varepsilon\right ) \right|^{2}
	+f_{Rh}\left (\varepsilon\right)\sum_{r=3}^{4}\left|\sum_{n}e^{in\omega t}G^{r}_{11,lr;-n0}
	\left ( \varepsilon\right )\right|^{2}\right]\right\}.
	\end{split}
	\label{Eq22}
\end{equation}
\end{widetext}	
We emphasize the last two subscripts in Eq. (\ref{Eq22}) refer to Fourier indices. Next, time-averaged current or dc current over one period $T$ ($T=2\pi/\omega$)
can also be written as Floquet-Landauer formalism \cite{Kohler,Rudner,Cuevas}:
\begin{equation}
	\begin{split}
	&\langle I_L(t) \rangle=\frac{1}{T}\int_{0}^{T} I \left(t\right) dt\\
	&=\sum_{n}\frac{e}{h} \int d\varepsilon \left\{\left ( T^{\left ( n \right )}_{ReLe}
	\left ( \varepsilon  \right )+T^{\left ( n \right )}_{RhLe}\left ( \varepsilon  \right )
	+T^{\left ( n \right )}_{LhLe}\left ( \varepsilon \right ) \right )
	f_{Le}\left(\varepsilon \right)\right.\\
	&\left.-T^{\left ( n \right )}_{LeLh}\left( \varepsilon\right)f_{Lh}(\varepsilon)
	-T^{\left ( n \right )}_{LeRe}\left( \varepsilon\right) f_{Re}\left(\varepsilon \right)-T^{\left ( n \right )}_{LeRh}\left( \varepsilon\right)f_{Rh}(\varepsilon ) \right\}.
	\end{split}
	\label{Eq23}
	\end{equation}
The transmission coefficients are calculated by the retarded Green function
	\begin{equation}
	\begin{split}
	T_{\alpha \beta,\alpha^{'}\beta^{'}}^{\left(n\right)}\left(\varepsilon\right)=Tr\left[ \mathbf{\Gamma}^{\alpha}_{\beta}\mathbf{G}^{r,\beta\beta'}_{n0}\left(\varepsilon\right)\mathbf{\Gamma}^{\alpha'}_{\beta'}\mathbf{G}^{a,\beta'\beta}_{0n}\left(\varepsilon\right)\right].
	\end{split}
	\label{Eq24}
	\end{equation}
Here $\alpha$,$\alpha^{'}$ denote $L$,$R$ normal leads and $\beta$,$\beta^{'}$ denote the electron e or hole h part. The resulting time-averaged differential conductance $G(V)=\frac{d \langle I_L(t) \rangle}{d V} $ in the low temperature limit is
\begin{equation}
G \left ( V \right )=\frac{e^2}{h}\sum_{n}
\left[T_{ReLe}^{\left ( n \right )}\left ( V \right )+2T_{LhLe}^{\left ( n \right )}\left ( V \right )+T_{RhLe}^{\left ( n \right )}\left ( V \right )\right]
\label{Eq25}
\end{equation}
where $T_{ReLe}^{\left ( n \right )} \left( V\right) $,
$T_{LhLe}^{\left( n\right) }\left(V\right)$,
$T_{RhLe}^{\left ( n \right )} \left( V\right) $ denote normal transmission,
local Andreev reflection and crossed Andreev reflection coefficients
for incident electrons from the left lead
by absorbing (emission) $\left| n\right| $ photons if $n>0$ ($n<0$)
and outgoing electron (or hole) to the left or right leads. It can be evaluated that Eq. (\ref{Eq23}) and Eq. (\ref{Eq25}). is consistent with results in the Ref \cite{Danon} once returning to static limit ($A=0$). Additionally, the derivation of current for periodically-driven single impurity is analogical, except that $R$ lead is replaced
by superconductor terminal $S$.

\subsection*{B. Numerical results for anomalous photon-assisted
tunneling}
With the help of Floquet Green's function and iterative method [See Appendix A], we can numerically calculate the time-averaged conductances. Here
we take the low temperature limits
and set $\widetilde{\Delta}=220\mu eV $, $\alpha_{R}=0.28 eV  \AA $
and effective mass $m^{*}=0.02m_{e}$
($m_{e}$ is the mass of electron) \cite{Mourik,Danon}.
The lattice constant $a$ =3.5nm and
the resulting nearest hopping energy is $t_{N}=
\hbar^2/(2m^*a^2) \approx 156 meV$.
The coupling is
$\widetilde{ \Gamma}^{L}=\widetilde{\Gamma}^{R}=0.05t_{N}$
for the 1D Majorana nanowire.
For the single impurity, we set $E_{0}=0$ and
$\widetilde{\Gamma}^{sc}=\widetilde{\Gamma}^{L}=0.05$ in the unit of $\hbar \omega$.

\begin{figure}
	\includegraphics[width=1\columnwidth]{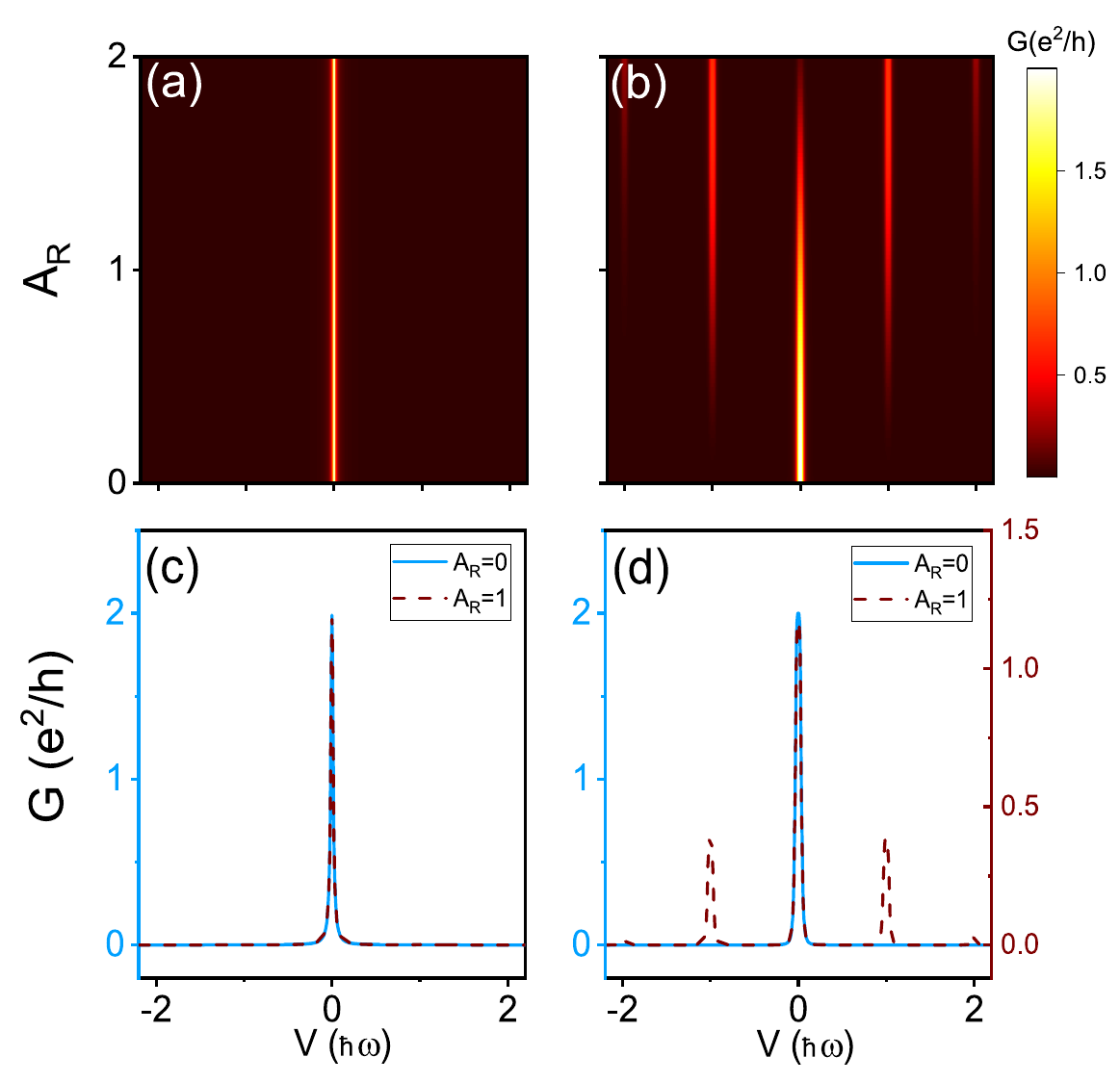}
	\centering
	\caption{ (a)-(d)
	The time-averaged conductances for the MBSs (a)(c)
	and single impurity (b)(d) driven by the harmonic magnetic field
	as functions of $A_{R}$ and bias $V$.
	Parameters are the same as Figs. 1(c)-(f).}
	\label{FIG2}
	\end{figure}

In Figs.~\ref{FIG1}(c-f), time-averaged conductances $G$ are shown
as functions of relative field strength
$A_{R}=\frac{A}{\hbar \omega}$ and the bias for the microwave-driven MBSs and single impurity. Figs.~\ref{FIG1}(e-f) are the cut-off of Figs.~\ref{FIG1}(c-d).
When driving field is absent ($A_R=0$),
both trivial ABSs and MBSs contribute a ZBCP. As $A_{R}$ climbs, the height of ZBCP for single impurity changes
dramatically and PAT sideband peaks emerges obviously
at harmonics $V=0,\pm 1\hbar \omega,\pm 2\hbar \omega,\cdots$,
indicating the absorption and emission of photons [Figs.~\ref{FIG1}(d, f)]. But
the ZBCP in the conductance spectroscopy remains the same
and no PAT peaks appears in the case of MBSs,
see Figs.~\ref{FIG1}(c, e).
This means no interaction between MBSs with $Q=0$ and photons. In the case of applying the harmonic magnetic field $\begin{pmatrix}
	0 ,& 0 ,& A\cos\left ( \omega t \right )
\end{pmatrix}$, similar results are exhibited,
see Figs.~~\ref{FIG2}(a-d). PAT sideband peaks disappear for MBSs in Figs.~\ref{FIG2}(a, c) but arise for single impurity in Figs.~\ref{FIG2}(b, d). This supports our argument that $\zeta_{z}$ for MBS is also zero.

\begin{figure}
	\includegraphics[width=1\columnwidth]{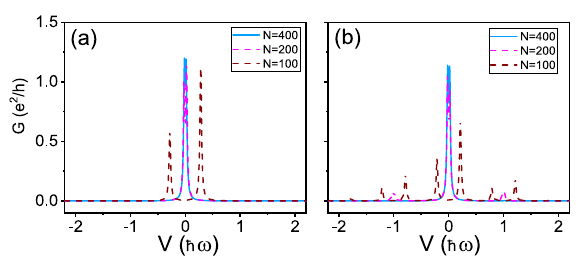}
	\centering 
	\caption{
 Time-averaged conductances $G$ vs. bias $V$
 for different nanowire lengths $N$ with $A_{R}=0$ (a) and $2$ (b).
 All the other parameters are the same as Fig.  \ref{FIG1}(e).}
	\label{FIG3}
\end{figure}

When Majorana nanowire length $N$ decreases,
two MBSs located at two opposite ends will overlap and
lead to an energy splitting.
In the meanwhile the lowest energy states $\psi_{0}$
formed by two MBSs will recover finite BCS charge $Q$ [as Eq.~(\ref{Eq14}) suggests].
In Figs.~\ref{FIG3}(a, b), we show the time-averaged conductances
for different nanowire lengths when $A_{R}=0$ and $A_{R}=2$.
Due to the enhanced hybridization between two MBSs,
the ZBCP splits into two peaks as the nanowire becomes short.
The stronger splitting is accompanied by a greater asymmetry
of electron-hole components
and more obvious PAT side peaks.
This anomalous PAT signal is consistent with our preceding analysis.
Disappearance of PAT signal actually reflects that a MBS remains unperturbed by local electromagnetic noise, which is crucial to implement a topological qubit. 

Besides the cases for MBSs and single impurity shown in Fig. \ref{FIG1},  
the real situation may be more complicated since MBSs and near-zero energy ABSs possibly coexist in the nanowire. 
At this time, PAT may still emerge due to the near-zero energy ABSs, 
even if MBSs are well-separated. 
MBSs' information here is inevitably interfered by ABSs. 
However, from the perspective of the application, 
the coexistence of MBSs and near-zero energy ABSs 
may not serve to prepare for topological qubits
since the probe is hard to be assured to selectively couple to one single MBS. 
Although MBSs cannot be well confirmed in this case, 
it has little application potential. What the experiment should pay attention for applications should be the case where PAT disappears.

\section{\label{sec4} Measurement of BCS charge}
\subsection*{A. Analytic form of time-averaged conductance}
Based on the projected Hamiltonian in Eq. (\ref{Eq8}), we are able to analyze how BCS information enters into time-averaged conductance. Take microwave-driving field as an example, the resulting projected self-energy terms of normal leads in the lowest energy states space are $\mathbf{\Sigma}^{r,p}=\mathbf{\Sigma}_{L}^{r,p}+\mathbf{\Sigma}_{R}^{r,p}$ with
\begin{equation}
\mathbf{\Sigma}_{\alpha=L,R}^{r,p}=\begin{pmatrix}
	-i\frac{\gamma_{\alpha}}{2}  & -i\frac{\xi_{\alpha}}{2}^{*} \\ -i\frac{\xi_{\alpha}}{2}
	& -i\frac{\gamma_{\alpha}}{2}
\end{pmatrix},
\label{Eq26}
\end{equation}
where $\gamma_{\alpha}=\tilde{\Gamma}^{\alpha} n_{\alpha}$,
$n_{L}=\sum_{s}\left(\left | u_{1s}^{(0)} \right |^{2}
+\left | v_{1s}^{(0)} \right |^{2}\right)$,
$n_{R}=\sum_{s} \left(\left | u_{Ns}^{(0)} \right |^{2}
+\left | v_{Ns}^{(0)} \right |^{2}\right)$,
$\xi_{L}=\tilde{\Gamma}^{L} \sum_{s}2u_{1s}^{(0)}v_{1s}^{(0)}$,
$\xi_{R}=\tilde{\Gamma}^{R}\sum_{s}2u_{Ns}^{(0)}v_{Ns}^{(0)}$.
We also introduce $\gamma_{Le}
=\sum_{s}\tilde{\Gamma}^{L}\left |u _{1 s}^{(0)} \right |^{2}$,
$\gamma_{Re}=\sum_{s}\tilde{\Gamma}^{R}\left |u _{Ns}^{(0)} \right |^{2}$,
$\gamma_{Lh}=\sum_{s}\tilde{\Gamma}^{L}\left |v _{1s}^{(0)} \right |^{2}$,
$\gamma_{Rh}=\sum_{s}\tilde{\Gamma}^{R}\left |v _{Ns}^{(0)} \right |^{2}$.
The linewidth functions are
\begin{equation}
	\mathbf{\Gamma}^{\alpha,p}=i\left( \mathbf{\Sigma}^{r,p}_{\alpha}-\mathbf{\Sigma}^{a,p}_{\alpha}\right) =\mathbf{\Gamma}^{\alpha,p}_{e}+\mathbf{\Gamma}^{\alpha,p}_{h}=\begin{pmatrix}
		\gamma_{\alpha}& \xi_{\alpha}^{*} \\ \xi_{\alpha}
		& \gamma_{\alpha}
		\end{pmatrix}
		\label{Eq27}
\end{equation}
Starting from Eq. (\ref{Eq8}) and Eq. (\ref{Eq26}), we can use the Dyson equation introduced in Ref.\cite{Sun1} to obtain the  analytic expressions for differential conductances (See details in the Appendix B).
Under the weak coupling strength
($\tilde{\Gamma}^{\alpha} \rightarrow 0$) and low temperature limits,
the $G\left(V\right)$ at the harmonics
$V=\varepsilon_{0}+n\hbar \omega$ $\left( \varepsilon_{0} \neq 0 , n=0,\pm 1, \cdots \right) $ denoted as $G_{n}$ can analytically be obtained as
\begin{equation}
G_{n}\approx\frac{e^2}{h}\frac{4\gamma_{Lh}}{\gamma^{2}}
(\gamma_{R}+2\gamma_{Le})J_{n}^{2}\left(QA_{R}\right),
\label{Eq28}
\end{equation}
where $\gamma=\gamma_L+\gamma_R$ and
$J_{n}\left(QA_{R}\right) $ is the nth-order Bessel
function of argument $QA_{R}$.
Close to $V=-\varepsilon_{0}+n\hbar \omega$,
the conductances are same as Eq. (\ref{Eq28}) except for exchanging $e$ and $h$.
Eq. (\ref{Eq28}) clearly suggest that the height of PAT peaks
are relevant to the BCS charge of the subgap bound state.
It indicates that we can extract the BCS charge
from the ratio of peak heights:
\begin{equation}
G_{1}/G_{0}={J_{1}^{2}\left(QA_{R}\right)}/{J_{0}^{2}\left(QA_{R}\right)}.
\label{Eq29}
\end{equation}
It is worth mentioning that here we use a local conductance
to detect BCS charge.
Compared with nonlocal conductances,
local conductance will give a stronger signal \cite{Danon}.

\subsection*{B. BCS charge extraction from the conductance spectroscopy}
\begin{figure}
	\includegraphics[width=1\columnwidth]{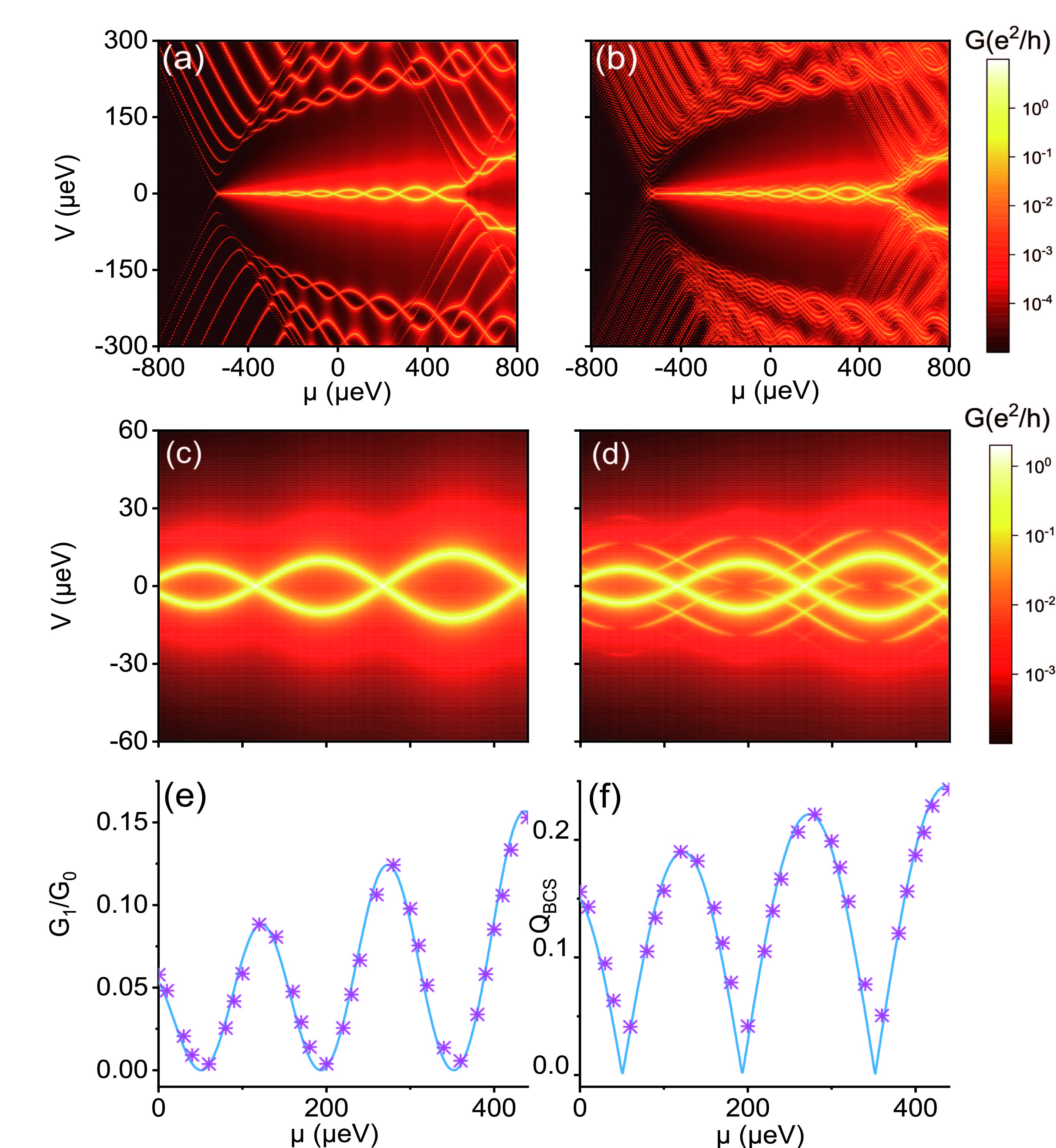}
	\centering 
	\caption{(a-d)
Time-averaged conductances $G$ of
microwave-driven Majorana nanowire as functions of $\mu$
and bias $V$ with $A_{R}=0$ (a, c) and $3$ (b, d).
(c, d) are the enlarged views of the low energy part in (a, b).
(e) The comparison between extracted ${G_{1}}/{G_{0}}$ (pink stars) from (d)
and ${J^{2}_{1}(QA_{R})}/{J^{2}_{0}(QA_{R})}$ (blue line)
with $Q$ obtained from $H_{nw}$ diagonalization.
(f) The comparison between $Q_{BCS}$ from conductance ratios ${G_{1}}/{G_{0}}$ (pink stars)
and `actual' $Q_{BCS}$ from direct diagonalization of $H_{nw}$ (blue line).
The parameters $N=400$, $V_{0}=600\mu eV$, and
$\hbar \omega =10\mu$eV. }
	\label{FIG4}
\end{figure}

\begin{figure}
	\includegraphics[width=1\columnwidth]{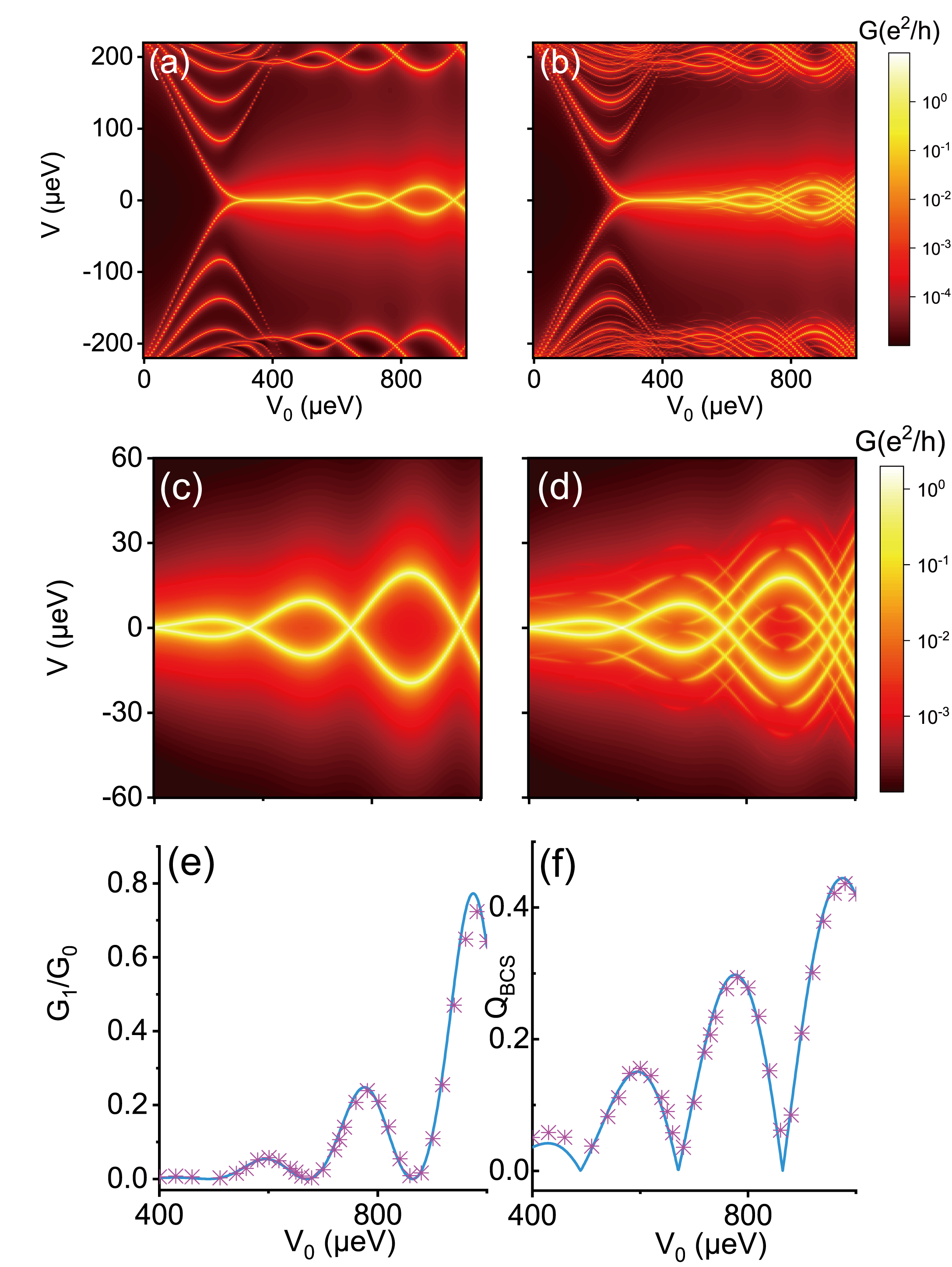}
	\centering
	\caption{ (a)-(d) Time-averaged conductances $G$ of
	microwave-driven Majorana nanowire as functions of static Zeeman field $V_{0}$ and bias $V$ at $\mu=0$ when $A_{R}=0$ (a, c) and $A_{R}=3$ (b, d).
	(c) and (d) are the enlarged views of (a) and (b).
	(e) The comparison between $G_{1}/G_{0}$ (pink stars) extracted
	from (d) and $J_{1}^{2}(QA_{R})/J_{0}^{2}(QA_{R})$ (blue line)
	where $Q$ is just BCS charge obtained from Hamiltonian $H_{nw}$ diagonalization.
	(f) The comparison between $Q_{BCS}$ obtained from conductances ratios $G_{1}/G_{0}$ (pink stars)
	and `actual' $Q_{BCS}$ from direct diagonalizing $H_{nw}$.
	The other parameters are the same as Fig.~\ref{FIG4}.}
	\label{FIG5}
\end{figure}

We try to numerically explore the time-averaged conductance as a function of
chemical potential $\mu$ (see Fig.~\ref{FIG4}) and Zeeman field $V_{0}$ (see Fig.~\ref{FIG5}) for microwave-driven Majorana nanowire. In Fig.~\ref{FIG4},
without the driving field ($A_R=0$),
the energy gap closure occurs at $\mu =\mu_c \approx \pm 550\mu eV $, which is roughly the topological transition point $\mu_c=\pm \sqrt{V_{0}^{2}-\widetilde{\Delta}^{2}}$ \cite{Lutchyn,Oreg}.
The ZBCP appears, and the nanowire is in the topological phase with a pair of MBSs
when $|\mu| < \mu_c$.
With the increase of $\mu$ from $-\mu_c$,
the ZBCP is almost unaffected at low $\mu$, but gradually split
and the lowest energy states $\varepsilon_0$ present
the characteristics of typical Majorana oscillation
[see Figs.~\ref{FIG4}(a,c)] \cite{Das2}.
The enhancement of the split (i.e. $\varepsilon_0$)
originates from the increase of coupling strength between MBSs
when $\mu$ raises.
After applying microwave field ($A_R = 3$),
the conductance spectroscopy become complicated [Fig.~\ref{FIG4}(b)].
Massive PAT sideband peaks concentrate on the region beyond induced topological gap
where electron-like and hole-like quasi-particle states stay. Near ZBCP, the photonic sideband peaks appear to be more obvious as $\mu$ grows.
From the enlarged view only concerning of
the lowest energy state in Fig.~\ref{FIG4}(d),
these sideband peaks can be seen clearly,
as a result of the coupling of MBSs leading to a nonzero BCS charge $Q$.
In Fig.~\ref{FIG4}(e),
we can extract a series of conductance peak ratios $G_{1}/G_{0}$
of zeroth and 1st harmonics from Fig.~\ref{FIG4}(d), and also show
${J_{1}^{2}\left(QA_{R}\right)}/{J_{0}^{2}\left(QA_{R}\right)}$ with
$Q$ obtained by numerically diagonalizing
isolated nanowire Hamiltonian $H_{nw}$.
There is a strong correlation between ${G_{1}}/{G_{0}}$ [pink stars] and ${J_{1}^{2}\left(QA_{R}\right)}/{J_{0}^{2}\left(QA_{R}\right)}$ [blue line].
Furthermore, in Fig.~\ref{FIG4}(f), we use these extracted conductance ratios ${G_{1}}/{G_{0}}$
to fit Eq. (\ref{Eq29}) to obtain $Q_{BCS}$ [pink stars]. It is found to be well consistent with the `actual' $Q_{BCS}$ [blue line]
from $H_{nm}$ diagonalization
\footnote{ The fitting for square ratio of Bessel function may be multivalued. But we can avoid this ambiguity by considering a small field strength and $Q<1$ }.
Comparing Figs.~\ref{FIG4}(d) and \ref{FIG4}(f), BCS charge and energy splitting oscillate out of phase due to $Q$ is equal to $\frac{d\varepsilon_0}{d\mu}$ \cite{Danon}.

Similar results are shown by varying the static Zeeman field $V_0$ in Fig.~\ref{FIG5}. With $A_{R}=0$, a gap closure and ZBCP appears at $V_{0} \approx 300\mu eV$ because of topological transition in Fig.~\ref{FIG5}(a). Then, as $V_{0}$ grows, a Majorana oscillation can be also observed as Zeeman field also modulates MBSs' coupling strength [See Figs.~\ref{FIG5}(a)(c)]. When microwave field exists ($A_{R} = 3$), PAT sideband peaks emerges clearly in Figs.~\ref{FIG5}(b)(d). We can repeat the previous process to compare $G_{1}/G_{0}$ and extract the BCS charge. The results still match the expectations [See Figs.~\ref{FIG5}(e)(f).]  In general, this provides a reliable way to detect BCS charges
by measuring the conductance ratios ${G_{1}}/{G_{0}}$.
In principle, if the driving field is locally applied on the wire,
the local BCS charge information can be also extracted.

\section{\label{sec5} Measurement of BCS spin components}
\begin{figure}
	\includegraphics[width=1\columnwidth]{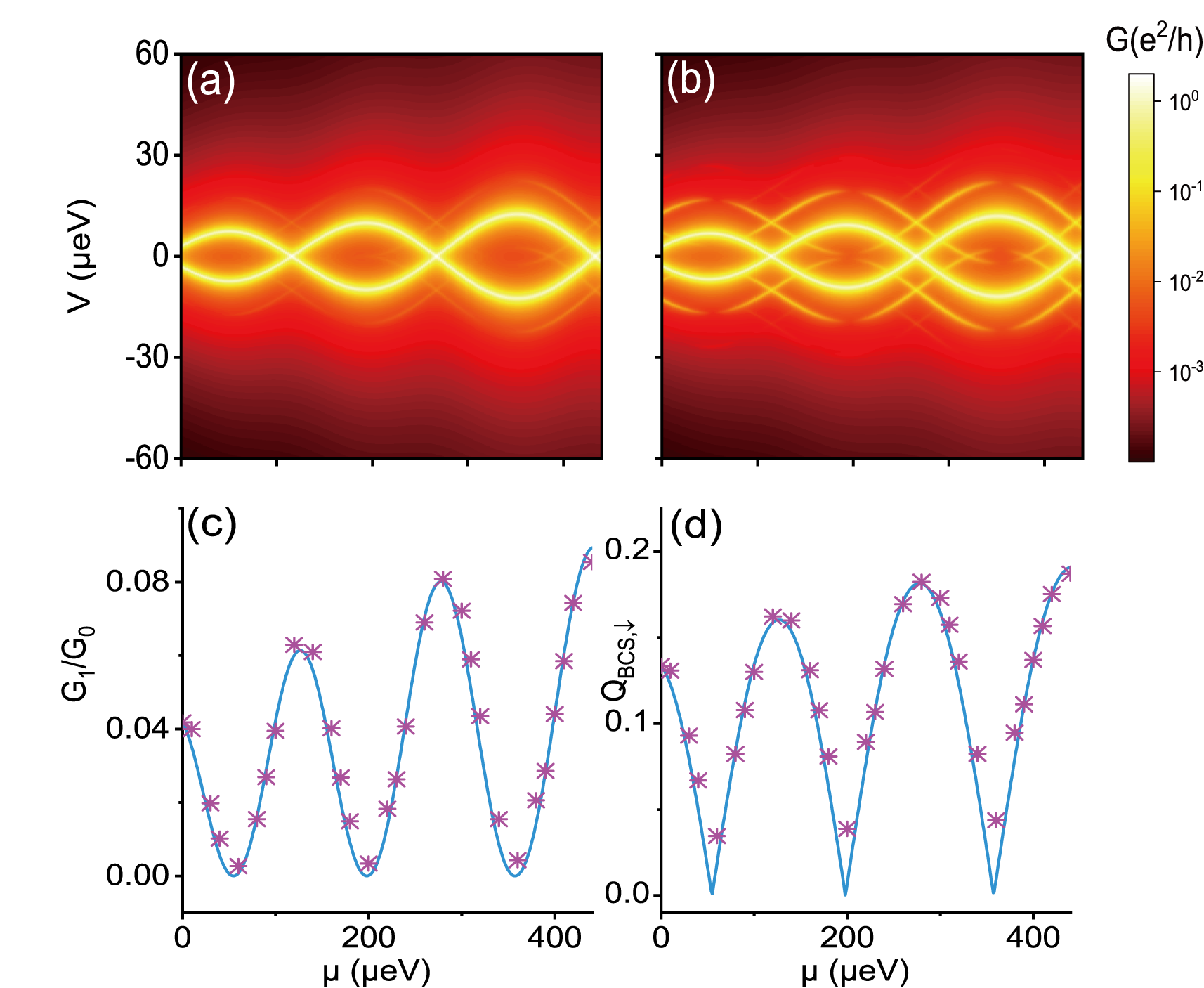}
	\centering 
	\caption{(a, b) Time-averaged conductance $G$ versus $\mu$ and $V$
at $V_{0}=600 \mu eV$ under the joint external field $H_{f1}$.
The initial phases of the two fields are $\phi_{1} =\phi_{2}=0$ (a)
and $\phi_{1} =0, \phi_{2}=\pi$ (b).
(c) The comparison between the extracted ${G_{1}}/{G_{0}}$ (pink stars) from (b)
and ${J^{2}_{1}(Q_{\downarrow} A_{R})}/{J^{2}_{0}(Q_{\downarrow} A_{R})}$ (blue line).
(d) The $Q_{BCS,\downarrow}$ obtained from the ${G_{1}}/{G_{0}}$ in (b) (pink stars)
and `actual' $Q_{BCS,\downarrow}$ by diagonalizing Hamiltonian $H_{nm}$ (blue line). Other Parameters are the same as Fig.~\ref{FIG4}.}
	\label{FIG6}
\end{figure}

Enlightened by the Eq. (\ref{Eq8}), we consider that the Majorana nanowire is driven
by both harmonic magnetic field along z direction and microwave field.
The external field Hamiltonian is
\begin{equation}
	\begin{split}
	H_{f1}&=H_{fm}+H_{fz}\\
	&=\sum_{i,s}\frac{A}{2}\cos\left (\omega t+\phi_{1}\right )c^{\dagger}_{is}c_{is}\\
	&+\sum_{i,s,s'} \frac{A}{2}\cos\left(\omega t+\phi_{2}\right)
	c^{\dagger}_{is}(\sigma^z)_{ss'} c_{is'}.
	\end{split}
	\label{Eq30}
\end{equation}

Here we have set the amplitude of two fields are equal,
which can be realized even without knowing $g$-factor
[see Appendix C].
When the two fields have the same initial phase $\phi_{1}=\phi_{2}=0$,
the renormalization factor in Eq. (\ref{Eq8})
is $Q_{\uparrow}=\sum_{i}\left(\left | u_{i\uparrow}^{(0)} \right |^{2}-\left|v_{i\uparrow}^{(0)}\right|^{2}\right)$.
But the renormalization factor is
$Q_{\downarrow}=\sum_{i}\left(\left | u_{i\downarrow}^{(0)} \right |^{2}-\left|v_{i\downarrow}^{(0)}\right|^{2}\right)$
when two initial phases have $\pi$ difference (e.g. $\phi_{2}=\pi$, $\phi_{1}=0$).
So we expect that $Q_{\uparrow}$ and $Q_{\downarrow}$,
the BCS spin components along the $z$ direction,
can be measured from conductance ratios under this joint field.

Figure~\ref{FIG6} shows time-averaged $G$ as a function of $\mu$ under the joint field with $\phi_{1}=\phi_{2}=0$ (a) and $\phi_{1}=0,\phi_{2}=\pi$ (b), respectively. Here the PAT peaks also exhibit as expected. Moreover, they are much clearer in Fig.~\ref{FIG6}(b) than in Fig.~\ref{FIG6}(a). The reason is that BCS charge for overlapping MBSs has a spin polarization : $Q_{BCS, \downarrow} >> Q_{BCS, \uparrow}$. This conforms to the Majorana wavefunction has large spin $\downarrow$ component for a strong magnetic field. From Fig.~\ref{FIG6}(b), we directly extract $G_{1}/G_{0}$ [pink stars] and find it is well consistent with $J_{1}^{2}(Q_{\downarrow}A_{R})/J_{0}^{2}(Q_{\downarrow}A_{R})$ [blue line]
, see Fig.~\ref{FIG6}(c). Similarly, in Fig.~\ref{FIG6}(d), $Q_{BCS,\downarrow}$
can be obtained by fitting Eq.~(\ref{Eq29}) using the extracted
$G_{1}/G_{0}$ [pink stars], and is almost the same as the `actual' $Q_{BCS,\downarrow}$ [blue line]
by diagonalizing Hamiltonian $H_{nw}$ (Due to errors in numerical extraction process, the deviations may occur when $Q_{BCS} \rightarrow 0$ ).

\begin{figure}
	\includegraphics[width=1\columnwidth]{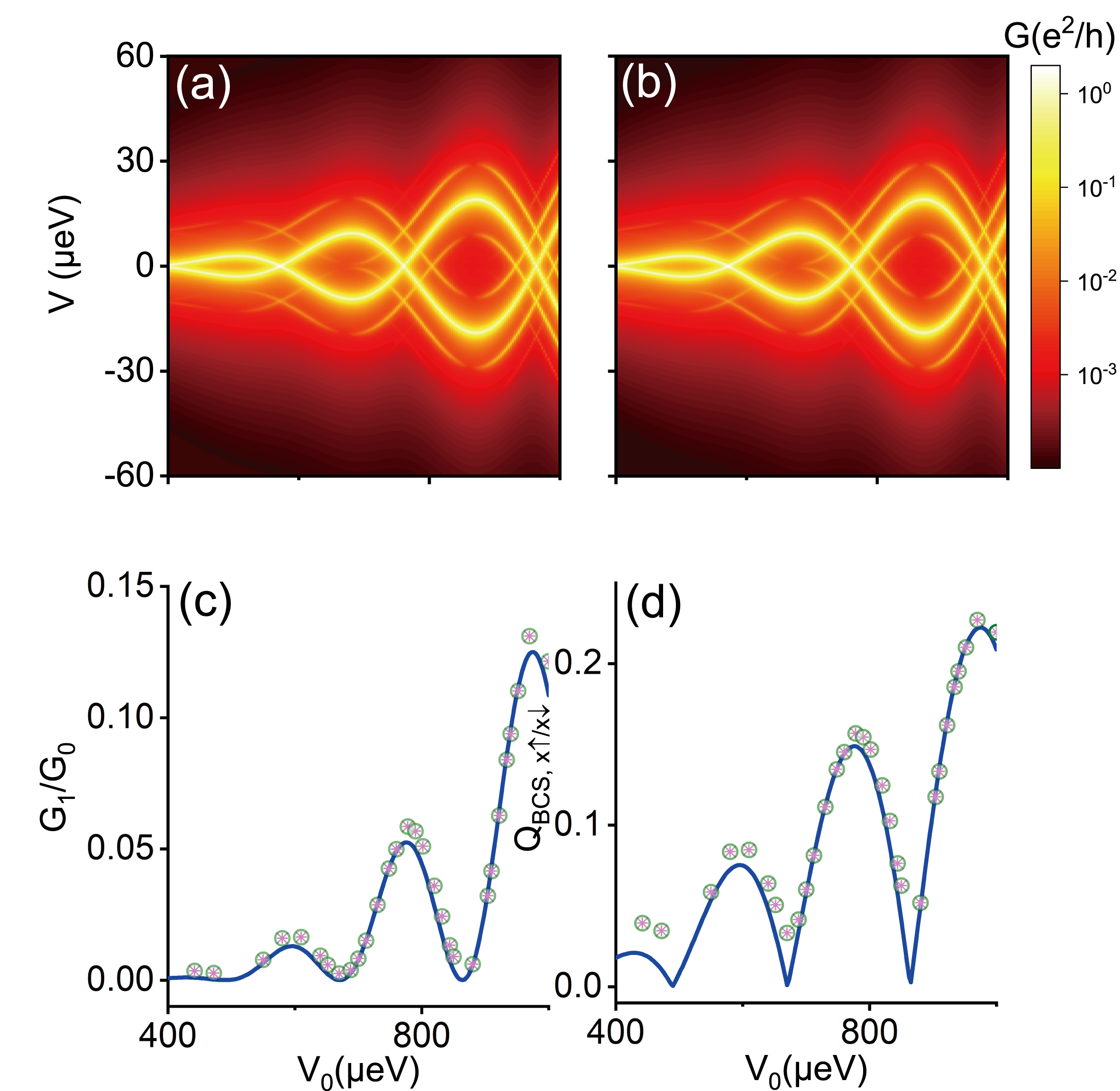}
	\centering 
	\caption{(a, b) Time-averaged conductance $G$ versus $V_{0}$ and $V$ at $\mu=0$ under the joint external field $H_{f2}$.
The initial phases of the two fields are $\phi_{1} =\phi_{2}=0$ (a)
and $\phi_{1} =0, \phi_{2}=\pi$ (b).
(c) The comparison between the extracted ${G_{1}}/{G_{0}}$ (pink stars for (a) and green circles for (b))
and ${J^{2}_{1}(QA_{R})}/{J^{2}_{0}(QA_{R})}$  where $Q=\sum_{i} \left(\left | u_{i,x\uparrow/x\downarrow}^{(0)} \right |^{2}
-\left | v_{i,x\uparrow/x\downarrow}^{(0)} \right |^{2} \right)$ (both of spin components are represented by dark blue line due to the coincidence).
(d) The $Q_{BCS}$ obtained from the ${G_{1}}/{G_{0}}$ [pink stars for (a) and green circles for (b)]
and `actual' $Q_{BCS}$ by diagonalizing Hamiltonian $H_{nm}$ (dark blue line for both spin components). Other Parameters are the same as Fig.~\ref{FIG4}.}
	\label{FIG7}
\end{figure}

BCS spin component along any direction can
be obtained by choosing appropriate harmonic magnetic field direction in principle. For example, when the Majorana nanowire is driven
by both harmonic magnetic field along x direction and microwave field.
The external field Hamiltonian is
\begin{equation}
	\begin{split}
	H_{f2}&=H_{fm}+H_{fx}\\
	&=\sum_{i,s}\frac{A}{2}\cos\left (\omega t+\phi_{1}\right )c^{\dagger}_{is}c_{is}\\
	&+\sum_{i,s,s'} \frac{A}{2}\cos\left(\omega t+\phi_{2}\right)
	c^{\dagger}_{is}(\sigma^x)_{ss'} c_{is'}.
	\end{split}
	\label{Eq31}
\end{equation}
Similarly, we can extract the BCS spin component $Q_{x\uparrow,x\downarrow}$ along x direction by adjusting phases $\phi_{1}=\phi_{2}=0$ or $\phi_{1}=0,\phi_{2}=\pi$, as shown in Figs.~\ref{FIG7}(a-d). Different from the Fig.~\ref{FIG6}, the conductance spectroscopies for $\phi_{1}=\phi_{2}=0$ in Fig.~\ref{FIG7}(a) and $\phi_{1}=0,\phi_{2}=\pi$ in Fig.~\ref{FIG7}(b) are totally the same, as a result of $Q_{BCS,x\uparrow}=Q_{BCS,x\downarrow}$. This equality origins from the symmetry of our model. It also means BCS spin polarization along x direction $\zeta_{x}=\sum_{i,s}(u_{is}u^{*}_{i\overline{s}}-v_{is}v^{*}_{i\overline{s}})$ is zero \cite{Serina} ($\overline{s}$ is the opposite direction of spin $s$). Besides, we also compare the conductance ratios $G_{1}/G_{0}$ and $Q_{BCS,x\uparrow,x\downarrow}$ with ${J^{2}_{1}(QA_{R})}/{J^{2}_{0}(QA_{R})}$ and `actual' $Q_{BCS,x\uparrow,x\downarrow}$ from $H_{nm}$ diagonalization. As shown in Figs.~\ref{FIG7}(c)(d), the comparison results are still consistent.

\section{\label{sec7} Summary and Discussion.}
In this paper, we show an anomalous PAT for the MBS where the sideband peaks are correlated with Majorana nonlocality in the periodically-driven Majorana nanowire. The PAT sideband peaks should disappear for well-separated MBSs but reappear for overlapping MBSs.  We also investigate the conductance spectroscopy by varying chemical potential and Zeeman field to study the PAT characteristics for Majorana oscillations. And a systematic scheme is further proposed to measure the BCS information of subgap states.

For the experiment realizations,
PAT in the periodically-driven Majorana nanowire is in principle attainable, especially in view that the research for PAT and periodically-driven system has developed as a mature field and covered a wide range as stated in the introduction. For instance, the interplay between microwave driving field and superconductors which we concern has produced many works in recent years such as the measurement of $4 \pi$ periodic Josephson effect \cite{Wiedenmann} and PAT into YSR states \cite{Peters}. One experiment very related to our setup is a recent work considering two coupled Majorana nanowires where one of them is applied by microwave field \cite{Zanten}. They investigate the PAT signals on the charge stability diagram and give a spectroscopic measurement of inter-pair coupling $E_{M}$ between zero modes on two superconducting islands. Our device is similar to them for one periodically-driven Majorana nanowire coupled to two normal leads 
and should be also achievable.

One potential question of our scheme in the experiment detection may be that thermal broadening could cover the sideband peaks. 
Then, the PAT signals in our previous results may be all invisible. 
The thermal broadening from finite temperature effect ignored in our calculation can be calculated as a convolution with the derivative of Fermi distribution. 
It will affect the line shape of conductance peaks in the same way, thus does not affect the conductances peaks ratios \cite{Danon}. In experiment, the temperature $T$ is usually tens of $mK$ corresponding to several $\mu eV $ for thermal broadening $k_{B}T$ ($k_{B}$ is Boltzmann constant). This is much lower than the superconducting gap which is a few hundreds of $\mu eV $ \cite{Mourik}. 
Therefore, by adjusting experiment parameters appropriately, 
the case satisfying ($\Gamma$, $k_{B}T$) (peaks broadening, several $\mu eV $) 
$ \ll \hbar \omega$ (photon energy, $\sim 10\mu eV$ ) $ \ll \widetilde{\Delta}$ (superconducting gap, several $100 \mu eV $ ) is possible in principle
to be realized to make sideband peaks distinct.

Although Majorana fermions are proved to exhibit an anomalous PAT in our work, it still remains a question whether this can be used to identify the emergence of topological MBSs in experiment, especially considering the various complex situations in the present experiments. This is, of course, not the main goal of our work. 
Here we could conclude some remarks. Overall, anomalous PAT sideband peaks have connection with Majorana nonlocality or BCS charge of subgap states. According to the Ref \cite{Pan}, the origin of ZBCP can be divided to three scenarios: good, bad and ugly. Only subgap states for good ZBCP case corresponds to a pair of topological MBSs (i.e. two MBSs located at two ends of the nanowire). For bad or ugly cases, the near-zero energy ABSs are composed of two partially or highly overlapped MBSs. Therefore, from their differences on Majorana nonlocality, PAT sideband peaks or BCS charge should exhibit distinguishable characteristics for good and bad/ugly cases within a range of parameters in principle. But some special cases, such as partially-separated ABSs composed of two MBSs separated by a distance of order of Majorana decay length \cite{Moore2,Stanescu} or quasi-Majoranas with two opposite spin polarizations \cite{Vuik}, may still need more detailed information. Additionally, the BCS spin information which can be extracted in our scheme is also helpful. For example, when quasi-MBSs and overlapping MBSs have a finite BCS charge, given their difference on spin polarizations, it is expected to extract BCS spin information and find  $Q_{BCS, \downarrow} \gg Q_{BCS, \uparrow}$ for topological MBSs (see Fig.~\ref{FIG6}) and $Q_{BCS, \downarrow} \approx Q_{BCS, \uparrow}$ for quasi-MBSs. In short, our scheme is at least able to provide some prejudgments for the existence of topological MBSs or for topological qubit implementation, in view of the close link between BCS charge and Majorana nonlocality.

\section*{Acknowledgments.}
This work was financially supported by National Key R and D Program of China (Grant No. 2017YFA0303301) and NSF-China (Grant No. 11921005) and the Strategic Priority Research Program of Chinese Academy of Sciences (Grant No.XDB28000000).
\\
\\

\appendix
\section{\label{A} Floquet Green's functions in Floquet extended space}
In this Appendix, we briefly introduce how to compute those Floquet Green's functions in Section III. For a more expansive and detailed descriptions, see Refs. \cite{Carlos, Liu2, Yang}.
We first write the  time-dependent Shr$\ddot{o}$dinger equation (in units of $\hbar=1$)
\begin{equation}
	i\partial_{t}\psi_{j} \left(t\right)= H\left(t\right) \psi_{j} \left(t\right),
	\label{EqA1}
\end{equation}
where $H(t+T)=H(t)$ under the periodically driving field. The wavefunction $\psi_{j}\left(t\right)$ can be decomposed into $\psi_{j}\left(t\right)=\sum_{m}e^{-i\epsilon_{j}t+im\omega t}\widetilde{\psi}^{m}_{j}$ in the Floquet extended space. Substituting this expression into the Eq.~(\ref{EqA1}) and multiplying on the left by $\frac{1}{T}\int_{0}^{T} dt e^{-in\omega t}$, we can get
\begin{equation}
H^{F}\widetilde{\psi}_{j}=\epsilon_{j}\widetilde{\psi}_{j},
\label{EqA2}
\end{equation}
where $\widetilde{\psi}_{j}$ is a vector in the Floquet extended space, $H^{F}$ is the Floquet Hamiltonian and its components can be defined as
\begin{equation}
	\begin{split}
H^{F}_{nm}&=\frac{1}{T}\int_{0}^{T}dte^{-in\omega t}\left(H\left(t\right)-i\partial_{t}\right)e^{im\omega t}\\
&=\frac{1}{T}\int_{0}^{T}dte^{i(m-n)\omega t}H\left(t\right)+m\omega \delta_{nm}.
    \end{split}
	\label{EqA3}
\end{equation}
Thus the time-dependent Hamiltonian in Eq.~(\ref{EqA1}) will be just changed into the time-independent Hamiltonian in Eq.~(\ref{EqA2}), by considering $m$ and $n$ index as Floquet basis. Similarly, we can give the retarded Floquet Green's function with the help of the Dyson equation in the Floquet extended space
\begin{equation}
	G^{r, -1}\left(E\right)=E*I-H^{F}-\sum_{\alpha=L,R}\Sigma^{r}_{\alpha},
\end{equation}
where $I$ is the identity matrix. The Floquet Hamiltonian for our system is
\begin{equation}
	\mathbf{H}^{F}=\begin{pmatrix}
	\cdots  & \mathbf{H}_{1} & \mathbf{0} & \mathbf{0} & \cdots \\
	\cdots & \mathbf{H}_{0}-\omega \mathbf{I}_{4N\times 4N} & \mathbf{H}_{1} & \mathbf{0} & \cdots \\
	\cdots & \mathbf{H}_{-1} & \mathbf{H}_{0} & \mathbf{H}_{1} &\cdots \\
	\cdots & \mathbf{0} & \mathbf{H}_{-1} & \mathbf{H}_{0} +\omega\mathbf{I}_{4N\times 4N} & \cdots \\
	\cdots & \mathbf{0} & \mathbf{0} & \mathbf{H}_{-1} & \cdots
	\end{pmatrix}.
	\label{EqA4}
	\end{equation}
where $N$ is the number of lattice sites for nanowire. $H_{0}$ is just the nanowire Hamiltonian $H_{nm}$. The off-diagonal elements $H_{1,-1}$ is related to the periodically driving field: $\mathbf{H}_{1}=\left( \mathbf{H}_{-1}\right) ^{\dagger}=\frac{A}{2}\mathbf{\tau}^{z}I_{2N \times 2N}$ for microwave field and $\frac{A}{2}\mathbf{\sigma}^{z}\mathbf{\tau}^{z}I_{N \times N}$ for harmonic magnetic field along z direction. The $\Sigma^{r}_{\alpha}$ is the self-energy from normal leads $\alpha$ and is related to the linewidth function matrix

\begin{equation}
\begin{split}
	\mathbf{\Gamma}^{L,R}&=i\left ( \mathbf{\Sigma} _{L,R}^{r}-\mathbf{\Sigma}_{L,R}^{a} \right )_{4NM\times4NM}\\
	&=\begin{pmatrix}
    (\mathbf{\Gamma}^{L,R}_{e})_{2NM\times 2NM} & 0 \\ 0 & (\mathbf{\Gamma}^{L,R}_{h})_{2NM\times 2NM}
	\end{pmatrix},
	\label{S17}
\end{split}
	\end{equation}
with $M$ is the Floquet space dimension which is infinite in principle. $\Sigma^{r}=\left(\Sigma^{a}\right)^{\dagger}$. $(\mathbf{\Gamma}^{L}_{e,h})_{ij,lr;mn}=\widetilde{\Gamma} ^{L}\delta_{ij}\delta_{i1}\delta_{lr}\delta_{mn}$ and $(\mathbf{\Gamma}^{R}_{e,h})_{ij,lr;mn}=\widetilde{\Gamma} ^{R}\delta_{ij}\delta_{iN}\delta_{lr}\delta_{mn}$, where $l,r$ only take the value of $1,2$ for electron part and $3,4$ for hole part.
To conduct numerical calculations in our main text, we truncate the Floquet space dimension into $[-7,7]$ (M=15).
In addition, with the help of recursive Green's function equations:
\begin{equation}
\begin{split}
	G^{\left(i+1\right),r}_{i+1,i+1}\left(E\right)&=[g^{\left(0\right),-1}_{i+1,i+1}\left(E\right)-V_{i+1,i}G^{\left(i\right),r}_{i,i}\left(E\right)V_{i,i+1}]^{-1}\\
	G^{\left(i+1\right),r}_{1,i+1}\left(E\right)&=G^{\left(i\right),r}_{1,i}\left(E\right)V_{i,i+1}G^{\left(i+1\right),r}_{i+1,i+1}\left(E\right)
\end{split}
\label{EqA7}
\end{equation}
where $G^{r}$ is the dressed Green's function involve the couplings between the sites, $g^{\left(0\right)}$ is the isolated Green's function for one site, $V_{i,j}$ is the hopping energy between site i and site j, we can obtain the $G^{r}_{n,n}$ and $G^{,r}_{1,n}$ for any $n$ starting from $G^{\left(1\right),r}_{1,1}=g^{\left(0\right)}_{1,1}$. We also denote all these terms on sites are put in the Floquet $\bigotimes$ BdG space.
\\
\\

\begin{widetext}
\section{\label{B} Derivations of analytic form for the time-averaged conductances}
In this appendix, we try to analyze the Floquet Green's functions in the projected space $(\psi^{\dagger}_{0},\psi_{0})$.  Using the Dyson equation, we can express the retarded Green's function $G^{r,p}$ in the projected space as \cite{Sun1}
\begin{equation}
	\left\{\begin{array}{l}
		G^{r,p}_{11;mn}=g^{r,p}_{11;mn}+\sum_{n_{1},n_{2}}G_{11;mn_{1}}^{r,p}\left ( \Sigma _{11;n_{1}n_{1}}^{r,p}\delta_{n_{1},n_{2}}+\Sigma ^{r,p}_{12;n_{1}n_{1}}D_{22;n_{1}n_{2}}\Sigma ^{r,p}_{21;n_{2}n_{2}} \right )g_{11;n_{2}n}^{r,p}\\G^{r,p}_{22;mn}=g^{r,p}_{22;mn}+\sum_{n_{1},n_{2}}G_{22;mn_{1}}^{r,p}\left ( \Sigma _{22;n_{1}n_{1}}^{r,p}\delta_{n_{1},n_{2}}+\Sigma ^{r,p}_{21;n_{1}n_{1}}D_{11;n_{1}n_{2}}\Sigma^{r,p}_{12;n_{2}n_{2}} \right )g_{22;n_{2}n}^{r,p}
		\\G^{r,p}_{12;mn}=\sum_{n_{1}} G^{r,p}_{11;mn_{1}}\Sigma_{12;n_{1}n_{1}}^{r}D_{22;n_{1}n}
		\\G^{r,p}_{21;mn}=\sum_{n_{1}} G^{r,p}_{22;mn_{1}}\Sigma_{21;n_{1}n_{1}}^{r,p}D_{11;n_{1}n},
	\end{array}\right.
	\label{B1}
	\end{equation}
where $\Sigma^{r,p}_{mn}=\Sigma^{r,p}\delta_{nm}$ due to time-independence and $\Sigma^{r,p}$ corresponds exactly to Eq. (\ref{Eq26}). Here the first two indices denote the Nambu spinor in the projected space $(\psi^{\dagger}_{0},\psi_{0})$and the other two indices denote Fourier indices or Floquet indices. Based on Eq. (\ref{Eq8}), the bare Green function $g^{r,p}_{mn}$ without coupling to normal leads is derived as (in units of $\hbar=1$)
\begin{equation}
g^{r,p}_{mn}=\begin{pmatrix}
\sum_{k}\frac{J_{k+m}\left ( QA_{R} \right )J_{k+n}\left ( QA_{R} \right )}{\varepsilon-\varepsilon _{0}-k\omega+i0^{+}} & 0 \\ 0
&\sum_{k}\frac{J_{k-m}\left ( QA_{R} \right )J_{k-n}\left ( QA_{R} \right )}{\varepsilon+\varepsilon _{0}+k\omega+i0^{+}}
\end{pmatrix}.
\label{B2}
\end{equation}
Here $A_{R}=A/\omega$. And $D_{ii;mn}$ $\left(i=1,2\right)$ in Eq. (\ref{B1})
can be calculated using recursive method :
\begin{equation}
D_{ii;mn}=g^{r,p}_{ii;mn}+\sum_{m_{1}}g^{r,p}_{ii;mm_{1}}\Sigma^{r,p}_{ii;m_{1}m_{1}}g^{r,p}_{ii;m_{1}n} +\cdots.
\label{B3}
\end{equation}
In the approximation of $\omega \gg \gamma=\gamma_{L}+\gamma_{R}$ ($\gamma_{L,R}$ is defined in Sec. IV),
we can have taken the approximation \cite{Sun1}:
\begin{equation}
\sum_{n,m}\frac{f_{nm}}{\left(\varepsilon+\varepsilon_{0}+n\omega+i0^{+}\right)
\left(\varepsilon+\varepsilon_{0}+m\omega+i0^{+}\right) }
\approx \sum_{n}\frac{f_{nn}}{\left(\varepsilon+\varepsilon_{0}+n\omega+i0^{+}\right)^2 }
\label{S32}
\end{equation}
for any function $f_{nm}$.
Thus, the projected retarded Green functions of Eq.~(\ref{B1})
can be summarized as
\begin{equation}
\begin{split}
G^{r,p}_{11;mn}&=\sum_{k}\frac{J_{m+k}\left ( QA_{R} \right)J_{n+k}\left ( QA_{R} \right )}{\varepsilon-\varepsilon _{0}-k\omega+i\frac{\gamma }{2}+\sum_{l}\frac{J^{2}_{k+l}\left(2QA_{R}\right)\left |\xi\right |^{2}/4}{ \varepsilon+ \varepsilon_{0}+l\omega+i\frac{\gamma}{2}}}\\
G^{r,p}_{22;mn}&=\sum_{k}\frac{J_{k-m}\left ( QA_{R} \right)J_{k-n}\left ( QA_{R} \right )}{\varepsilon+\varepsilon _{0}+k\omega+i\frac{\gamma }{2}+\sum_{l}\frac{J^{2}_{k+l}\left(2QA_{R}\right)\left |\xi\right |^{2}/4}{ \varepsilon- \varepsilon_{0}-l\omega+i\frac{\gamma}{2}}}\\
G^{r,p}_{12;mn}&=\sum_{ks}\frac{J_{m+k}\left ( QA_{R} \right)J_{s-n}\left ( QA_{R} \right )J_{k+s}\left( 2QA_{R}\right) }{\left(  \varepsilon-\varepsilon _{0}-k\omega+i\frac{\gamma }{2}+\sum_{l}\frac{J^{2}_{k+l}\left(2QA_{R}\right)\left |\xi\right |^{2}/4}{ \varepsilon+ \varepsilon_{0}+l\omega+i\frac{\gamma}{2} }\right) \left( \varepsilon+ \varepsilon_{0}+s\omega+i\frac{\gamma}{2} \right) }\left( -i\xi^{*}/2\right) \\
G^{r,p}_{21;mn}&=\sum_{ks}\frac{J_{k-m}\left ( QA_{R} \right)J_{s+n}\left ( QA_{R} \right )J_{k+s}\left( 2QA_{R}\right) }{\left( \varepsilon+\varepsilon _{0}+k\omega+i\frac{\gamma }{2}+\sum_{l}\frac{J^{2}_{k+l}\left(2QA_{R}\right)\left |\xi\right |^{2}/4}{ \varepsilon-\varepsilon_{0}-l\omega+i\frac{\gamma}{2} }\right) \left( \varepsilon-\varepsilon_{0}-s\omega+i\frac{\gamma}{2} \right) }\left( -i\xi/2\right).
\end{split}
\label{B5}
\end{equation}
\end{widetext}
Here $\xi=\xi_{L}+\xi_{R}$ where $\xi_{L,R}$ is defined in Sec. IV. Now we focus on the region close to harmonics,
namely near $\varepsilon=\pm \varepsilon_{0}+ n \omega$ ($\varepsilon_{0} > 0, n=\cdots -1,0,1,2 \cdots$).
For convenience, we denote time-averaged conductances at these resonant levels
as $G_{n}$. Despite the lattice indices no longer existing due to projection operation on the lowest energy states, we can still substitute Green's functions in Eq. (\ref{B5}) along with linewidth functions in Eq. (\ref{Eq27}) into Eq. (\ref{Eq24}) and Eq. (\ref{Eq25}) to obtain transmission coefficients and the resulting conductance (all of these quantities are regarded as matrices in the projected space with Floquet indices). Considering
the weak coupling limit ($\gamma_{L}+\gamma_{R} \ll \omega$), the 
time-averaged differential conductance is simplified into,
\begin{equation}
\begin{split}
G(V) &\approx \frac{e^{2}}{h}\sum _{k}\left [ \left ( \gamma_{R}+2 \gamma_{Le}\right )\gamma_{Lh}\frac{J_{k}^{2}\left ( QA_{R} \right )}{\left ( V-\varepsilon_{0}-k\omega \right )^{2}+\frac{1}{4}\gamma ^{2}}\right.\\
& \left.+\left ( \gamma_{R}+2\gamma_{Lh} \right )\gamma_{Le}\frac{J_{k}^{2}\left ( QA_{R} \right )}{\left ( V+\varepsilon_{0}+k\omega \right )^{2}+\frac{1}{4}\gamma ^{2}} \right ].
\label{B6}
\end{split}
\end{equation}
When two MBSs are well-separated, $\gamma_{Lh}=\gamma_{Le}=\frac{1}{2}\gamma_{L}$,
$\varepsilon_{0}\rightarrow 0$. And at $V=n\omega$,
\begin{equation}
\begin{split}
G_{n}&\approx \frac{4e^{2}}{h}\frac{\gamma_{L}}{\gamma}J_{n}^{2}\left ( QA_{R} \right ).
\end{split}
\label{S35}
\end{equation}
If no driving field exists $A_{R}=0$, then $n=0$, we recover the result of quantized ZBCP $\frac{2e^{2}}{h}$ once $\gamma_{L}=\gamma_{R}$. When two MBSs are coupled, namely, $\varepsilon_{0} > 0$, we can obtain the
time-averaged conductance at $V=\varepsilon_{0}+n \omega$ is
\begin{equation}
G_{n}\approx \frac{e^{2}}{h}\frac{4\gamma_{Lh}}{\gamma^{2}}\left ( \gamma_{R}+2\gamma_{Le} \right )J_{n}^{2}\left ( QA_{R} \right ).
\label{S36}
\end{equation}
Similarly we find that at $V=-\varepsilon_{0}+n \omega$, the time-averaged conductance is
\begin{equation}
G_{n}\approx \frac{e^{2}}{h}\frac{4\gamma_{Le}}{\gamma^{2}}\left( \gamma_{R}+2\gamma_{Lh} \right )J_{n}^{2}\left ( QA_{R} \right ).
\label{S37}
\end{equation}
Here we could find the time-averaged differential conductance peaks at harmonics are directly proportional to $J_{n}^{2}\left( QA_{R}\right)$.

\section{\label{C} Determination of the position that two fields meet for MBS}
In experiment, the g-factor of the hybrid nanowires may be modified and sometimes difficult to be determined. Aimed to MBSs, due to their wavefunctions' spin polarization properties, we can still find a way to approximately locate the position that two fields are equal, even if without knowing g-factor. By this we can move forward to obtain g-factor and then the BCS spin components.

\begin{figure}
	\includegraphics[width=0.9\columnwidth]{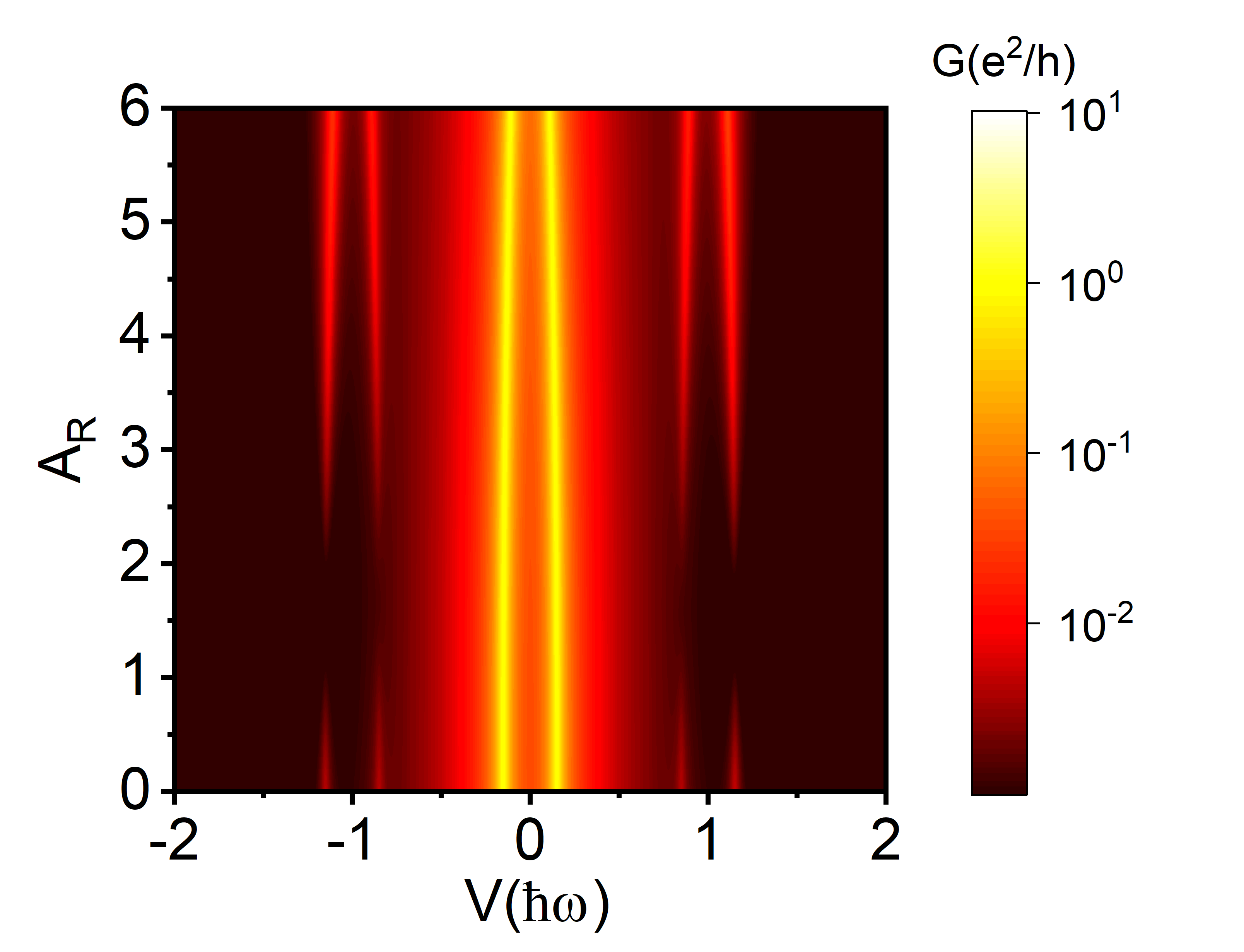}
	\centering
	\caption{The time-averaged conductance spectroscopy
	for weakly coupled MBSs under the joint field versus bias $V$ and relative microwave intensity $A_{R}=A/ \hbar \omega$. Here $ \hbar \omega=20\mu eV$, $V_{0}=600\mu eV$ and $\mu=0$.
	The other parameters are the same as Fig. \ref{FIG4}.}
	\label{FIG8}	
\end{figure}

In Fig.~\ref{FIG8} we focus on a pair of MBSs with finite coupling in a short nanowire, driven by the joint field
\begin{equation}
    \begin{split}
	H_{f}&=H_{fm}+H_{fz} \\
	&=\sum_{i,ss'}\left(\frac{A}{2}(\sigma^0)_{ss'}+\frac{V_{z}}{2}(\sigma^z)_{ss'}\right)\cos\left (\omega t\right)c^{\dagger}_{is}c_{is'}
    \end{split}
	\label{EqC1}
\end{equation}
The amplitude of Zeeman field is set as $V_{z}=2 \hbar \omega$.
We increase the amplitude of microwave field $A$ from zero.
Fig.~\ref{FIG8} shows the time-averaged conductance spectroscopy for coupled MBSs. There are two high main peaks at $V=\pm\varepsilon_0$
due to the coupling of two MBSs. The PAT sideband peaks emerge at harmonics $V=\pm\varepsilon_0 \pm \hbar\omega$.
With the increase of $A$ from zero, the PAT sideband peaks
show obviously a minimum region.
The minimum value of the PAT sideband peaks locate about at $A=V_{z}$.
So from the PAT conductance spectroscopy, we can ascertain the position
of $A=V_{z}=\frac{1}{2}g\mu_B B$ and further obtain the $g$-factor.
The reasons are as follows:
the joint driving field $H_{f}$ projected to the lowest energy
states space $(\psi_0^{\dagger}, \psi_0)$ will be
\begin{equation}
\begin{split}
H_f^p &=H_{fm}^p+H_{fz}^p \\
&=\left[\frac{A(Q_{\uparrow}+Q_{\downarrow})}{2}
 +\frac{V_{z}(Q_{\uparrow}-Q_{\downarrow})}{2}\right]\cos\left (\omega t\right) \nonumber\\
&= \left[\frac{(A+V_{z})Q_{\uparrow}}{2}
 +\frac{(A-V_{z})Q_{\downarrow}}{2}\right]\cos\left (\omega t\right).
 \label{EqC2}
\end{split}
 \end{equation}
 Here $H_f^p$ is the effective driving field for the lowest subgap states.
In addition, $ \left[\frac{(A+V_{z})Q_{\uparrow}}{2 \hbar \omega}+\frac{(A-V_{z})Q_{\downarrow}}{2\hbar \omega } \right]$ will also be
the argument of Bessel function, as Eq. (\ref{Eq28}) in the main text indicates. For MBSs, $|Q_{\uparrow} | $ is rather tinier than $ | Q_{\downarrow}|$.
As a result, the effective driving field approximately
is $\frac{(A-V_{z})Q_{\downarrow}}{2}\cos(\omega t)$.
In Fig.~\ref{FIG8}, when microwave field strength $A$ is zero,
BCS spin components $-V_z Q_{\downarrow}$ contributes to the external driving field.
When $A<V_{z}$, as $A$ grows, $|(A-V_{z})Q_{\downarrow}|$ decreases,
and the height of PAT sideband peaks also decline.
At $A=V_{z}$, $(A-V_{z})Q_{\downarrow}=0$, leading to a very weak PAT sideband peak.
Once $A$ goes across $V_{z}$, $|(A-V_{z})Q_{\downarrow}|$ increases from 0 once more
and PAT sideband peaks could go up again.
So PAT sideband peaks will reach around a minimum when $A=V_{z}$ for MBSs. In summary, in the topological MBSs system, by scanning the intensity of
the microwave field to find the position for the PAT sideband peaks' minimum, we can approximately determine the point where two fields meet and further extract both of BCS spin components. Moreover, this method could also help to estimate $g$-factor.
\bibliography{ref}

\begin{thebibliography}{58}%
\makeatletter
\providecommand \@ifxundefined [1]{%
 \@ifx{#1\undefined}
}%
\providecommand \@ifnum [1]{%
 \ifnum #1\expandafter \@firstoftwo
 \else \expandafter \@secondoftwo
 \fi
}%
\providecommand \@ifx [1]{%
 \ifx #1\expandafter \@firstoftwo
 \else \expandafter \@secondoftwo
 \fi
}%
\providecommand \natexlab [1]{#1}%
\providecommand \enquote  [1]{``#1''}%
\providecommand \bibnamefont  [1]{#1}%
\providecommand \bibfnamefont [1]{#1}%
\providecommand \citenamefont [1]{#1}%
\providecommand \href@noop [0]{\@secondoftwo}%
\providecommand \href [0]{\begingroup \@sanitize@url \@href}%
\providecommand \@href[1]{\@@startlink{#1}\@@href}%
\providecommand \@@href[1]{\endgroup#1\@@endlink}%
\providecommand \@sanitize@url [0]{\catcode `\\12\catcode `\$12\catcode
  `\&12\catcode `\#12\catcode `\^12\catcode `\_12\catcode `\%12\relax}%
\providecommand \@@startlink[1]{}%
\providecommand \@@endlink[0]{}%
\providecommand \url  [0]{\begingroup\@sanitize@url \@url }%
\providecommand \@url [1]{\endgroup\@href {#1}{\urlprefix }}%
\providecommand \urlprefix  [0]{URL }%
\providecommand \Eprint [0]{\href }%
\providecommand \doibase [0]{https://doi.org/}%
\providecommand \selectlanguage [0]{\@gobble}%
\providecommand \bibinfo  [0]{\@secondoftwo}%
\providecommand \bibfield  [0]{\@secondoftwo}%
\providecommand \translation [1]{[#1]}%
\providecommand \BibitemOpen [0]{}%
\providecommand \bibitemStop [0]{}%
\providecommand \bibitemNoStop [0]{.\EOS\space}%
\providecommand \EOS [0]{\spacefactor3000\relax}%
\providecommand \BibitemShut  [1]{\csname bibitem#1\endcsname}%
\let\auto@bib@innerbib\@empty
\bibitem [{\citenamefont {Wilczek}(2009)}]{Wilczek}%
  \BibitemOpen
  \bibfield  {author} {\bibinfo {author} {\bibfnamefont {F.}~\bibnamefont
  {Wilczek}},\ }\href@noop {} {\bibfield  {journal} {\bibinfo  {journal} {Nat.
  Phys.}\ }\textbf {\bibinfo {volume} {5}},\ \bibinfo {pages} {614} (\bibinfo
  {year} {2009})}\BibitemShut {NoStop}%
\bibitem [{\citenamefont {Kitaev}(2001)}]{Kitaev}%
  \BibitemOpen
  \bibfield  {author} {\bibinfo {author} {\bibfnamefont {A.~Y.}\ \bibnamefont
  {Kitaev}},\ }\href@noop {} {\bibfield  {journal} {\bibinfo  {journal} {Phys.
  Usp.}\ }\textbf {\bibinfo {volume} {44}},\ \bibinfo {pages} {131} (\bibinfo
  {year} {2001})}\BibitemShut {NoStop}%
\bibitem [{\citenamefont {Alicea}(2012)}]{Alicea}%
  \BibitemOpen
  \bibfield  {author} {\bibinfo {author} {\bibfnamefont {J.}~\bibnamefont
  {Alicea}},\ }\href@noop {} {\bibfield  {journal} {\bibinfo  {journal} {Rep.
  Prog. Phys}\ }\textbf {\bibinfo {volume} {75}},\ \bibinfo {pages} {076501}
  (\bibinfo {year} {2012})}\BibitemShut {NoStop}%
\bibitem [{\citenamefont {Elliott}\ and\ \citenamefont
  {Franz}(2015)}]{Elliott}%
  \BibitemOpen
  \bibfield  {author} {\bibinfo {author} {\bibfnamefont {S.~R.}\ \bibnamefont
  {Elliott}}\ and\ \bibinfo {author} {\bibfnamefont {M.}~\bibnamefont
  {Franz}},\ }\href {https://doi.org/10.1103/RevModPhys.87.137} {\bibfield
  {journal} {\bibinfo  {journal} {Rev. Mod. Phys.}\ }\textbf {\bibinfo {volume}
  {87}},\ \bibinfo {pages} {137} (\bibinfo {year} {2015})}\BibitemShut
  {NoStop}%
\bibitem [{\citenamefont {Kitaev}(2003)}]{Kitaev2}%
  \BibitemOpen
  \bibfield  {author} {\bibinfo {author} {\bibfnamefont {A.}~\bibnamefont
  {Kitaev}},\ }\href@noop {} {\bibfield  {journal} {\bibinfo  {journal} {Ann.
  Phys.}\ }\textbf {\bibinfo {volume} {303}},\ \bibinfo {pages} {2} (\bibinfo
  {year} {2003})}\BibitemShut {NoStop}%
\bibitem [{\citenamefont {Nayak}\ \emph {et~al.}(2008)\citenamefont {Nayak},
  \citenamefont {Simon}, \citenamefont {Stern}, \citenamefont {Freedman},\ and\
  \citenamefont {Das~Sarma}}]{Nayak}%
  \BibitemOpen
  \bibfield  {author} {\bibinfo {author} {\bibfnamefont {C.}~\bibnamefont
  {Nayak}}, \bibinfo {author} {\bibfnamefont {S.~H.}\ \bibnamefont {Simon}},
  \bibinfo {author} {\bibfnamefont {A.}~\bibnamefont {Stern}}, \bibinfo
  {author} {\bibfnamefont {M.}~\bibnamefont {Freedman}},\ and\ \bibinfo
  {author} {\bibfnamefont {S.}~\bibnamefont {Das~Sarma}},\ }\href@noop {}
  {\bibfield  {journal} {\bibinfo  {journal} {Rev. Mod. Phys.}\ }\textbf
  {\bibinfo {volume} {80}},\ \bibinfo {pages} {1083} (\bibinfo {year}
  {2008})}\BibitemShut {NoStop}%
\bibitem [{\citenamefont {Aasen}\ \emph {et~al.}(2016)\citenamefont {Aasen},
  \citenamefont {Hell}, \citenamefont {Mishmash}, \citenamefont {Higginbotham},
  \citenamefont {Danon}, \citenamefont {Leijnse}, \citenamefont {Jespersen},
  \citenamefont {Folk}, \citenamefont {Marcus}, \citenamefont {Flensberg},\
  and\ \citenamefont {Alicea}}]{Aasen}%
  \BibitemOpen
  \bibfield  {author} {\bibinfo {author} {\bibfnamefont {D.}~\bibnamefont
  {Aasen}}, \bibinfo {author} {\bibfnamefont {M.}~\bibnamefont {Hell}},
  \bibinfo {author} {\bibfnamefont {R.~V.}\ \bibnamefont {Mishmash}}, \bibinfo
  {author} {\bibfnamefont {A.}~\bibnamefont {Higginbotham}}, \bibinfo {author}
  {\bibfnamefont {J.}~\bibnamefont {Danon}}, \bibinfo {author} {\bibfnamefont
  {M.}~\bibnamefont {Leijnse}}, \bibinfo {author} {\bibfnamefont {T.~S.}\
  \bibnamefont {Jespersen}}, \bibinfo {author} {\bibfnamefont {J.~A.}\
  \bibnamefont {Folk}}, \bibinfo {author} {\bibfnamefont {C.~M.}\ \bibnamefont
  {Marcus}}, \bibinfo {author} {\bibfnamefont {K.}~\bibnamefont {Flensberg}},\
  and\ \bibinfo {author} {\bibfnamefont {J.}~\bibnamefont {Alicea}},\
  }\href@noop {} {\bibfield  {journal} {\bibinfo  {journal} {Phys. Rev. X}\
  }\textbf {\bibinfo {volume} {6}},\ \bibinfo {pages} {031016} (\bibinfo {year}
  {2016})}\BibitemShut {NoStop}%
\bibitem [{\citenamefont {Lutchyn}\ \emph {et~al.}(2010)\citenamefont
  {Lutchyn}, \citenamefont {Sau},\ and\ \citenamefont {Das~Sarma}}]{Lutchyn}%
  \BibitemOpen
  \bibfield  {author} {\bibinfo {author} {\bibfnamefont {R.~M.}\ \bibnamefont
  {Lutchyn}}, \bibinfo {author} {\bibfnamefont {J.~D.}\ \bibnamefont {Sau}},\
  and\ \bibinfo {author} {\bibfnamefont {S.}~\bibnamefont {Das~Sarma}},\
  }\href@noop {} {\bibfield  {journal} {\bibinfo  {journal} {Phys. Rev. Lett.}\
  }\textbf {\bibinfo {volume} {105}},\ \bibinfo {pages} {077001} (\bibinfo
  {year} {2010})}\BibitemShut {NoStop}%
\bibitem [{\citenamefont {Oreg}\ \emph {et~al.}(2010)\citenamefont {Oreg},
  \citenamefont {Refael},\ and\ \citenamefont {von Oppen}}]{Oreg}%
  \BibitemOpen
  \bibfield  {author} {\bibinfo {author} {\bibfnamefont {Y.}~\bibnamefont
  {Oreg}}, \bibinfo {author} {\bibfnamefont {G.}~\bibnamefont {Refael}},\ and\
  \bibinfo {author} {\bibfnamefont {F.}~\bibnamefont {von Oppen}},\ }\href@noop
  {} {\bibfield  {journal} {\bibinfo  {journal} {Phys. Rev. Lett.}\ }\textbf
  {\bibinfo {volume} {105}},\ \bibinfo {pages} {177002} (\bibinfo {year}
  {2010})}\BibitemShut {NoStop}%
\bibitem [{\citenamefont {Mourik}\ \emph {et~al.}(2012)\citenamefont {Mourik},
  \citenamefont {Zuo}, \citenamefont {Frolov}, \citenamefont {Plissard},
  \citenamefont {Bakkers},\ and\ \citenamefont {Kouwenhoven}}]{Mourik}%
  \BibitemOpen
  \bibfield  {author} {\bibinfo {author} {\bibfnamefont {V.}~\bibnamefont
  {Mourik}}, \bibinfo {author} {\bibfnamefont {K.}~\bibnamefont {Zuo}},
  \bibinfo {author} {\bibfnamefont {S.~M.}\ \bibnamefont {Frolov}}, \bibinfo
  {author} {\bibfnamefont {S.~R.}\ \bibnamefont {Plissard}}, \bibinfo {author}
  {\bibfnamefont {E.~P. A.~M.}\ \bibnamefont {Bakkers}},\ and\ \bibinfo
  {author} {\bibfnamefont {L.~P.}\ \bibnamefont {Kouwenhoven}},\ }\href@noop {}
  {\bibfield  {journal} {\bibinfo  {journal} {Science}\ }\textbf {\bibinfo
  {volume} {336}},\ \bibinfo {pages} {1003} (\bibinfo {year}
  {2012})}\BibitemShut {NoStop}%
\bibitem [{\citenamefont {Law}\ \emph {et~al.}(2009)\citenamefont {Law},
  \citenamefont {Lee},\ and\ \citenamefont {Ng}}]{Law}%
  \BibitemOpen
  \bibfield  {author} {\bibinfo {author} {\bibfnamefont {K.~T.}\ \bibnamefont
  {Law}}, \bibinfo {author} {\bibfnamefont {P.~A.}\ \bibnamefont {Lee}},\ and\
  \bibinfo {author} {\bibfnamefont {T.~K.}\ \bibnamefont {Ng}},\ }\href@noop {}
  {\bibfield  {journal} {\bibinfo  {journal} {Phys. Rev. Lett.}\ }\textbf
  {\bibinfo {volume} {103}},\ \bibinfo {pages} {237001} (\bibinfo {year}
  {2009})}\BibitemShut {NoStop}%
\bibitem [{\citenamefont {Fu}\ and\ \citenamefont {Kane}(2009)}]{Fu}%
  \BibitemOpen
  \bibfield  {author} {\bibinfo {author} {\bibfnamefont {L.}~\bibnamefont
  {Fu}}\ and\ \bibinfo {author} {\bibfnamefont {C.~L.}\ \bibnamefont {Kane}},\
  }\href@noop {} {\bibfield  {journal} {\bibinfo  {journal} {Phys. Rev. B}\
  }\textbf {\bibinfo {volume} {79}},\ \bibinfo {pages} {161408(R)} (\bibinfo
  {year} {2009})}\BibitemShut {NoStop}%
\bibitem [{\citenamefont {He}\ \emph {et~al.}(2014)\citenamefont {He},
  \citenamefont {Ng}, \citenamefont {Lee},\ and\ \citenamefont {Law}}]{He}%
  \BibitemOpen
  \bibfield  {author} {\bibinfo {author} {\bibfnamefont {J.~J.}\ \bibnamefont
  {He}}, \bibinfo {author} {\bibfnamefont {T.~K.}\ \bibnamefont {Ng}}, \bibinfo
  {author} {\bibfnamefont {P.~A.}\ \bibnamefont {Lee}},\ and\ \bibinfo {author}
  {\bibfnamefont {K.~T.}\ \bibnamefont {Law}},\ }\href@noop {} {\bibfield
  {journal} {\bibinfo  {journal} {Phys. Rev. Lett.}\ }\textbf {\bibinfo
  {volume} {112}},\ \bibinfo {pages} {037001} (\bibinfo {year}
  {2014})}\BibitemShut {NoStop}%
\bibitem [{\citenamefont {Pawlak}\ \emph {et~al.}(2019)\citenamefont {Pawlak},
  \citenamefont {Hoffman}, \citenamefont {Klinovaja}, \citenamefont {Loss},\
  and\ \citenamefont {Meyer}}]{Pawlak}%
  \BibitemOpen
  \bibfield  {author} {\bibinfo {author} {\bibfnamefont {R.}~\bibnamefont
  {Pawlak}}, \bibinfo {author} {\bibfnamefont {S.}~\bibnamefont {Hoffman}},
  \bibinfo {author} {\bibfnamefont {J.}~\bibnamefont {Klinovaja}}, \bibinfo
  {author} {\bibfnamefont {D.}~\bibnamefont {Loss}},\ and\ \bibinfo {author}
  {\bibfnamefont {E.}~\bibnamefont {Meyer}},\ }\href@noop {} {\bibfield
  {journal} {\bibinfo  {journal} {Prog. Part. Nucl. Phys.}\ }\textbf {\bibinfo
  {volume} {107}},\ \bibinfo {pages} {1} (\bibinfo {year} {2019})}\BibitemShut
  {NoStop}%
\bibitem [{\citenamefont {Lin}\ \emph {et~al.}(2012)\citenamefont {Lin},
  \citenamefont {Sau},\ and\ \citenamefont {Das~Sarma}}]{Lin}%
  \BibitemOpen
  \bibfield  {author} {\bibinfo {author} {\bibfnamefont {C.-H.}\ \bibnamefont
  {Lin}}, \bibinfo {author} {\bibfnamefont {J.~D.}\ \bibnamefont {Sau}},\ and\
  \bibinfo {author} {\bibfnamefont {S.}~\bibnamefont {Das~Sarma}},\ }\href@noop
  {} {\bibfield  {journal} {\bibinfo  {journal} {Phys. Rev. B}\ }\textbf
  {\bibinfo {volume} {86}},\ \bibinfo {pages} {224511} (\bibinfo {year}
  {2012})}\BibitemShut {NoStop}%
\bibitem [{\citenamefont {Ben-Shach}\ \emph {et~al.}(2015)\citenamefont
  {Ben-Shach}, \citenamefont {Haim}, \citenamefont {Appelbaum}, \citenamefont
  {Oreg}, \citenamefont {Yacoby},\ and\ \citenamefont {Halperin}}]{Ben}%
  \BibitemOpen
  \bibfield  {author} {\bibinfo {author} {\bibfnamefont {G.}~\bibnamefont
  {Ben-Shach}}, \bibinfo {author} {\bibfnamefont {A.}~\bibnamefont {Haim}},
  \bibinfo {author} {\bibfnamefont {I.}~\bibnamefont {Appelbaum}}, \bibinfo
  {author} {\bibfnamefont {Y.}~\bibnamefont {Oreg}}, \bibinfo {author}
  {\bibfnamefont {A.}~\bibnamefont {Yacoby}},\ and\ \bibinfo {author}
  {\bibfnamefont {B.~I.}\ \bibnamefont {Halperin}},\ }\href@noop {} {\bibfield
  {journal} {\bibinfo  {journal} {Phys. Rev. B}\ }\textbf {\bibinfo {volume}
  {91}},\ \bibinfo {pages} {045403} (\bibinfo {year} {2015})}\BibitemShut
  {NoStop}%
\bibitem [{\citenamefont {Domínguez}\ \emph {et~al.}(2017)\citenamefont
  {Domínguez}, \citenamefont {Cayao}, \citenamefont {San-Jose}, \citenamefont
  {Aguado}, \citenamefont {Yeyati},\ and\ \citenamefont {Prada}}]{Fernando}%
  \BibitemOpen
  \bibfield  {author} {\bibinfo {author} {\bibfnamefont {F.}~\bibnamefont
  {Domínguez}}, \bibinfo {author} {\bibfnamefont {J.}~\bibnamefont {Cayao}},
  \bibinfo {author} {\bibfnamefont {P.}~\bibnamefont {San-Jose}}, \bibinfo
  {author} {\bibfnamefont {R.}~\bibnamefont {Aguado}}, \bibinfo {author}
  {\bibfnamefont {A.~L.}\ \bibnamefont {Yeyati}},\ and\ \bibinfo {author}
  {\bibfnamefont {E.}~\bibnamefont {Prada}},\ }\href@noop {} {\bibfield
  {journal} {\bibinfo  {journal} {npj Quantum Mater}\ }\textbf {\bibinfo
  {volume} {2}},\ \bibinfo {pages} {13} (\bibinfo {year} {2017})}\BibitemShut
  {NoStop}%
\bibitem [{\citenamefont {Prada}\ \emph {et~al.}(2017)\citenamefont {Prada},
  \citenamefont {Aguado},\ and\ \citenamefont {San-Jose}}]{Prada1}%
  \BibitemOpen
  \bibfield  {author} {\bibinfo {author} {\bibfnamefont {E.}~\bibnamefont
  {Prada}}, \bibinfo {author} {\bibfnamefont {R.}~\bibnamefont {Aguado}},\ and\
  \bibinfo {author} {\bibfnamefont {P.}~\bibnamefont {San-Jose}},\ }\href@noop
  {} {\bibfield  {journal} {\bibinfo  {journal} {Phys. Rev. B}\ }\textbf
  {\bibinfo {volume} {96}},\ \bibinfo {pages} {085418} (\bibinfo {year}
  {2017})}\BibitemShut {NoStop}%
\bibitem [{\citenamefont {Deng}\ \emph {et~al.}(2018)\citenamefont {Deng},
  \citenamefont {Vaitiek\ifmmode~\dot{e}\else \.{e}\fi{}nas}, \citenamefont
  {Prada}, \citenamefont {San-Jose}, \citenamefont {Nyg\aa{}rd}, \citenamefont
  {Krogstrup}, \citenamefont {Aguado},\ and\ \citenamefont {Marcus}}]{Deng}%
  \BibitemOpen
  \bibfield  {author} {\bibinfo {author} {\bibfnamefont {M.-T.}\ \bibnamefont
  {Deng}}, \bibinfo {author} {\bibfnamefont {S.}~\bibnamefont
  {Vaitiek\ifmmode~\dot{e}\else \.{e}\fi{}nas}}, \bibinfo {author}
  {\bibfnamefont {E.}~\bibnamefont {Prada}}, \bibinfo {author} {\bibfnamefont
  {P.}~\bibnamefont {San-Jose}}, \bibinfo {author} {\bibfnamefont
  {J.}~\bibnamefont {Nyg\aa{}rd}}, \bibinfo {author} {\bibfnamefont
  {P.}~\bibnamefont {Krogstrup}}, \bibinfo {author} {\bibfnamefont
  {R.}~\bibnamefont {Aguado}},\ and\ \bibinfo {author} {\bibfnamefont {C.~M.}\
  \bibnamefont {Marcus}},\ }\href@noop {} {\bibfield  {journal} {\bibinfo
  {journal} {Phys. Rev. B}\ }\textbf {\bibinfo {volume} {98}},\ \bibinfo
  {pages} {085125} (\bibinfo {year} {2018})}\BibitemShut {NoStop}%
\bibitem [{\citenamefont {Pe\~naranda}\ \emph {et~al.}(2018)\citenamefont
  {Pe\~naranda}, \citenamefont {Aguado}, \citenamefont {San-Jose},\ and\
  \citenamefont {Prada}}]{Penaranda}%
  \BibitemOpen
  \bibfield  {author} {\bibinfo {author} {\bibfnamefont {F.}~\bibnamefont
  {Pe\~naranda}}, \bibinfo {author} {\bibfnamefont {R.}~\bibnamefont {Aguado}},
  \bibinfo {author} {\bibfnamefont {P.}~\bibnamefont {San-Jose}},\ and\
  \bibinfo {author} {\bibfnamefont {E.}~\bibnamefont {Prada}},\ }\href@noop {}
  {\bibfield  {journal} {\bibinfo  {journal} {Phys. Rev. B}\ }\textbf {\bibinfo
  {volume} {98}},\ \bibinfo {pages} {235406} (\bibinfo {year}
  {2018})}\BibitemShut {NoStop}%
\bibitem [{\citenamefont {Danon}\ \emph {et~al.}(2020)\citenamefont {Danon},
  \citenamefont {Hellenes}, \citenamefont {Hansen}, \citenamefont {Casparis},
  \citenamefont {Higginbotham},\ and\ \citenamefont {Flensberg}}]{Danon}%
  \BibitemOpen
  \bibfield  {author} {\bibinfo {author} {\bibfnamefont {J.}~\bibnamefont
  {Danon}}, \bibinfo {author} {\bibfnamefont {A.~B.}\ \bibnamefont {Hellenes}},
  \bibinfo {author} {\bibfnamefont {E.~B.}\ \bibnamefont {Hansen}}, \bibinfo
  {author} {\bibfnamefont {L.}~\bibnamefont {Casparis}}, \bibinfo {author}
  {\bibfnamefont {A.~P.}\ \bibnamefont {Higginbotham}},\ and\ \bibinfo {author}
  {\bibfnamefont {K.}~\bibnamefont {Flensberg}},\ }\href@noop {} {\bibfield
  {journal} {\bibinfo  {journal} {Phys. Rev. Lett.}\ }\textbf {\bibinfo
  {volume} {124}},\ \bibinfo {pages} {036801} (\bibinfo {year}
  {2020})}\BibitemShut {NoStop}%
\bibitem [{\citenamefont {M\'enard}\ \emph {et~al.}(2020)\citenamefont
  {M\'enard}, \citenamefont {Anselmetti}, \citenamefont {Martinez},
  \citenamefont {Puglia}, \citenamefont {Malinowski}, \citenamefont {Lee},
  \citenamefont {Choi}, \citenamefont {Pendharkar}, \citenamefont
  {Palmstr\o{}m}, \citenamefont {Flensberg}, \citenamefont {Marcus},
  \citenamefont {Casparis},\ and\ \citenamefont {Higginbotham}}]{Menard}%
  \BibitemOpen
  \bibfield  {author} {\bibinfo {author} {\bibfnamefont {G.~C.}\ \bibnamefont
  {M\'enard}}, \bibinfo {author} {\bibfnamefont {G.~L.~R.}\ \bibnamefont
  {Anselmetti}}, \bibinfo {author} {\bibfnamefont {E.~A.}\ \bibnamefont
  {Martinez}}, \bibinfo {author} {\bibfnamefont {D.}~\bibnamefont {Puglia}},
  \bibinfo {author} {\bibfnamefont {F.~K.}\ \bibnamefont {Malinowski}},
  \bibinfo {author} {\bibfnamefont {J.~S.}\ \bibnamefont {Lee}}, \bibinfo
  {author} {\bibfnamefont {S.}~\bibnamefont {Choi}}, \bibinfo {author}
  {\bibfnamefont {M.}~\bibnamefont {Pendharkar}}, \bibinfo {author}
  {\bibfnamefont {C.~J.}\ \bibnamefont {Palmstr\o{}m}}, \bibinfo {author}
  {\bibfnamefont {K.}~\bibnamefont {Flensberg}}, \bibinfo {author}
  {\bibfnamefont {C.~M.}\ \bibnamefont {Marcus}}, \bibinfo {author}
  {\bibfnamefont {L.}~\bibnamefont {Casparis}},\ and\ \bibinfo {author}
  {\bibfnamefont {A.~P.}\ \bibnamefont {Higginbotham}},\ }\href@noop {}
  {\bibfield  {journal} {\bibinfo  {journal} {Phys. Rev. Lett.}\ }\textbf
  {\bibinfo {volume} {124}},\ \bibinfo {pages} {036802} (\bibinfo {year}
  {2020})}\BibitemShut {NoStop}%
\bibitem [{\citenamefont {Tucker}\ and\ \citenamefont
  {Feldman}(1985)}]{Tucker}%
  \BibitemOpen
  \bibfield  {author} {\bibinfo {author} {\bibfnamefont {J.~R.}\ \bibnamefont
  {Tucker}}\ and\ \bibinfo {author} {\bibfnamefont {M.~J.}\ \bibnamefont
  {Feldman}},\ }\href@noop {} {\bibfield  {journal} {\bibinfo  {journal} {Rev.
  Mod. Phys.}\ }\textbf {\bibinfo {volume} {57}},\ \bibinfo {pages} {1055}
  (\bibinfo {year} {1985})}\BibitemShut {NoStop}%
\bibitem [{\citenamefont {Platero}\ and\ \citenamefont
  {Aguado}(2004)}]{Platero}%
  \BibitemOpen
  \bibfield  {author} {\bibinfo {author} {\bibfnamefont {G.}~\bibnamefont
  {Platero}}\ and\ \bibinfo {author} {\bibfnamefont {R.}~\bibnamefont
  {Aguado}},\ }\href@noop {} {\bibfield  {journal} {\bibinfo  {journal} {Phys.
  Rep.}\ }\textbf {\bibinfo {volume} {395}},\ \bibinfo {pages} {1} (\bibinfo
  {year} {2004})}\BibitemShut {NoStop}%
\bibitem [{\citenamefont {Tien}\ and\ \citenamefont {Gordon}(1963)}]{Tien}%
  \BibitemOpen
  \bibfield  {author} {\bibinfo {author} {\bibfnamefont {P.~K.}\ \bibnamefont
  {Tien}}\ and\ \bibinfo {author} {\bibfnamefont {J.~P.}\ \bibnamefont
  {Gordon}},\ }\href@noop {} {\bibfield  {journal} {\bibinfo  {journal} {Phys.
  Rev.}\ }\textbf {\bibinfo {volume} {129}},\ \bibinfo {pages} {647} (\bibinfo
  {year} {1963})}\BibitemShut {NoStop}%
\bibitem [{\citenamefont {Kot}\ \emph {et~al.}(2020)\citenamefont {Kot},
  \citenamefont {Drost}, \citenamefont {Uhl}, \citenamefont {Ankerhold},
  \citenamefont {Cuevas},\ and\ \citenamefont {Ast}}]{Kot}%
  \BibitemOpen
  \bibfield  {author} {\bibinfo {author} {\bibfnamefont {P.}~\bibnamefont
  {Kot}}, \bibinfo {author} {\bibfnamefont {R.}~\bibnamefont {Drost}}, \bibinfo
  {author} {\bibfnamefont {M.}~\bibnamefont {Uhl}}, \bibinfo {author}
  {\bibfnamefont {J.}~\bibnamefont {Ankerhold}}, \bibinfo {author}
  {\bibfnamefont {J.~C.}\ \bibnamefont {Cuevas}},\ and\ \bibinfo {author}
  {\bibfnamefont {C.~R.}\ \bibnamefont {Ast}},\ }\href@noop {} {\bibfield
  {journal} {\bibinfo  {journal} {Phys. Rev. B}\ }\textbf {\bibinfo {volume}
  {101}},\ \bibinfo {pages} {134507} (\bibinfo {year} {2020})}\BibitemShut
  {NoStop}%
\bibitem [{\citenamefont {Kouwenhoven}\ \emph {et~al.}(1994)\citenamefont
  {Kouwenhoven}, \citenamefont {Jauhar}, \citenamefont {McCormick},
  \citenamefont {Dixon}, \citenamefont {McEuen}, \citenamefont {Nazarov},
  \citenamefont {van~der Vaart},\ and\ \citenamefont {Foxon}}]{Kouwenhoven}%
  \BibitemOpen
  \bibfield  {author} {\bibinfo {author} {\bibfnamefont {L.~P.}\ \bibnamefont
  {Kouwenhoven}}, \bibinfo {author} {\bibfnamefont {S.}~\bibnamefont {Jauhar}},
  \bibinfo {author} {\bibfnamefont {K.}~\bibnamefont {McCormick}}, \bibinfo
  {author} {\bibfnamefont {D.}~\bibnamefont {Dixon}}, \bibinfo {author}
  {\bibfnamefont {P.~L.}\ \bibnamefont {McEuen}}, \bibinfo {author}
  {\bibfnamefont {Y.~V.}\ \bibnamefont {Nazarov}}, \bibinfo {author}
  {\bibfnamefont {N.~C.}\ \bibnamefont {van~der Vaart}},\ and\ \bibinfo
  {author} {\bibfnamefont {C.~T.}\ \bibnamefont {Foxon}},\ }\href@noop {}
  {\bibfield  {journal} {\bibinfo  {journal} {Phys. Rev. B}\ }\textbf {\bibinfo
  {volume} {50}},\ \bibinfo {pages} {2019} (\bibinfo {year}
  {1994})}\BibitemShut {NoStop}%
\bibitem [{\citenamefont {Sun}\ \emph {et~al.}(1998)\citenamefont {Sun},
  \citenamefont {Wang},\ and\ \citenamefont {Lin}}]{Sun3}%
  \BibitemOpen
  \bibfield  {author} {\bibinfo {author} {\bibfnamefont {Q.-f.}\ \bibnamefont
  {Sun}}, \bibinfo {author} {\bibfnamefont {J.}~\bibnamefont {Wang}},\ and\
  \bibinfo {author} {\bibfnamefont {T.-h.}\ \bibnamefont {Lin}},\ }\href@noop
  {} {\bibfield  {journal} {\bibinfo  {journal} {Phys. Rev. B}\ }\textbf
  {\bibinfo {volume} {58}},\ \bibinfo {pages} {13007} (\bibinfo {year}
  {1998})}\BibitemShut {NoStop}%
\bibitem [{\citenamefont {Keay}\ \emph {et~al.}(1995)\citenamefont {Keay},
  \citenamefont {Allen}, \citenamefont {Gal\'an}, \citenamefont {Kaminski},
  \citenamefont {Campman}, \citenamefont {Gossard}, \citenamefont
  {Bhattacharya},\ and\ \citenamefont {Rodwell}}]{Keay}%
  \BibitemOpen
  \bibfield  {author} {\bibinfo {author} {\bibfnamefont {B.~J.}\ \bibnamefont
  {Keay}}, \bibinfo {author} {\bibfnamefont {S.~J.}\ \bibnamefont {Allen}},
  \bibinfo {author} {\bibfnamefont {J.}~\bibnamefont {Gal\'an}}, \bibinfo
  {author} {\bibfnamefont {J.~P.}\ \bibnamefont {Kaminski}}, \bibinfo {author}
  {\bibfnamefont {K.~L.}\ \bibnamefont {Campman}}, \bibinfo {author}
  {\bibfnamefont {A.~C.}\ \bibnamefont {Gossard}}, \bibinfo {author}
  {\bibfnamefont {U.}~\bibnamefont {Bhattacharya}},\ and\ \bibinfo {author}
  {\bibfnamefont {M.~J.~W.}\ \bibnamefont {Rodwell}},\ }\href@noop {}
  {\bibfield  {journal} {\bibinfo  {journal} {Phys. Rev. Lett.}\ }\textbf
  {\bibinfo {volume} {75}},\ \bibinfo {pages} {4098} (\bibinfo {year}
  {1995})}\BibitemShut {NoStop}%
\bibitem [{\citenamefont {Mani}\ \emph {et~al.}(2002)\citenamefont {Mani},
  \citenamefont {Smet}, \citenamefont {von Klitzing}, \citenamefont
  {Narayanamurti}, \citenamefont {Johnson},\ and\ \citenamefont
  {Umansky}}]{Mani}%
  \BibitemOpen
  \bibfield  {author} {\bibinfo {author} {\bibfnamefont {R.~G.}\ \bibnamefont
  {Mani}}, \bibinfo {author} {\bibfnamefont {J.~H.}\ \bibnamefont {Smet}},
  \bibinfo {author} {\bibfnamefont {K.}~\bibnamefont {von Klitzing}}, \bibinfo
  {author} {\bibfnamefont {V.}~\bibnamefont {Narayanamurti}}, \bibinfo {author}
  {\bibfnamefont {W.~B.}\ \bibnamefont {Johnson}},\ and\ \bibinfo {author}
  {\bibfnamefont {V.}~\bibnamefont {Umansky}},\ }\href@noop {} {\bibfield
  {journal} {\bibinfo  {journal} {Nature (London)}\ }\textbf {\bibinfo {volume}
  {420}},\ \bibinfo {pages} {646} (\bibinfo {year} {2002})}\BibitemShut
  {NoStop}%
\bibitem [{\citenamefont {Zudov}\ \emph {et~al.}(2003)\citenamefont {Zudov},
  \citenamefont {Du}, \citenamefont {Pfeiffer},\ and\ \citenamefont
  {West}}]{Zudov}%
  \BibitemOpen
  \bibfield  {author} {\bibinfo {author} {\bibfnamefont {M.~A.}\ \bibnamefont
  {Zudov}}, \bibinfo {author} {\bibfnamefont {R.~R.}\ \bibnamefont {Du}},
  \bibinfo {author} {\bibfnamefont {L.~N.}\ \bibnamefont {Pfeiffer}},\ and\
  \bibinfo {author} {\bibfnamefont {K.~W.}\ \bibnamefont {West}},\ }\href@noop
  {} {\bibfield  {journal} {\bibinfo  {journal} {Phys. Rev. Lett.}\ }\textbf
  {\bibinfo {volume} {90}},\ \bibinfo {pages} {046807} (\bibinfo {year}
  {2003})}\BibitemShut {NoStop}%
\bibitem [{\citenamefont {Shi}\ and\ \citenamefont {Xie}(2003)}]{Shi}%
  \BibitemOpen
  \bibfield  {author} {\bibinfo {author} {\bibfnamefont {J.}~\bibnamefont
  {Shi}}\ and\ \bibinfo {author} {\bibfnamefont {X.~C.}\ \bibnamefont {Xie}},\
  }\href@noop {} {\bibfield  {journal} {\bibinfo  {journal} {Phys. Rev. Lett.}\
  }\textbf {\bibinfo {volume} {91}},\ \bibinfo {pages} {086801} (\bibinfo
  {year} {2003})}\BibitemShut {NoStop}%
\bibitem [{\citenamefont {Tang}\ \emph {et~al.}(2015)\citenamefont {Tang},
  \citenamefont {Zhang},\ and\ \citenamefont {Liu}}]{Tang}%
  \BibitemOpen
  \bibfield  {author} {\bibinfo {author} {\bibfnamefont {H.-Z.}\ \bibnamefont
  {Tang}}, \bibinfo {author} {\bibfnamefont {Y.-T.}\ \bibnamefont {Zhang}},\
  and\ \bibinfo {author} {\bibfnamefont {J.-J.}\ \bibnamefont {Liu}},\
  }\href@noop {} {\bibfield  {journal} {\bibinfo  {journal} {AIP Adv}\ }\textbf
  {\bibinfo {volume} {5}},\ \bibinfo {pages} {127129} (\bibinfo {year}
  {2015})}\BibitemShut {NoStop}%
\bibitem [{\citenamefont {Gonz\'alez}\ \emph {et~al.}(2020)\citenamefont
  {Gonz\'alez}, \citenamefont {Melischek}, \citenamefont {Peters},
  \citenamefont {Flensberg}, \citenamefont {Franke},\ and\ \citenamefont {von
  Oppen}}]{Gonzalez}%
  \BibitemOpen
  \bibfield  {author} {\bibinfo {author} {\bibfnamefont {S.~A.}\ \bibnamefont
  {Gonz\'alez}}, \bibinfo {author} {\bibfnamefont {L.}~\bibnamefont
  {Melischek}}, \bibinfo {author} {\bibfnamefont {O.}~\bibnamefont {Peters}},
  \bibinfo {author} {\bibfnamefont {K.}~\bibnamefont {Flensberg}}, \bibinfo
  {author} {\bibfnamefont {K.~J.}\ \bibnamefont {Franke}},\ and\ \bibinfo
  {author} {\bibfnamefont {F.}~\bibnamefont {von Oppen}},\ }\href@noop {}
  {\bibfield  {journal} {\bibinfo  {journal} {Phys. Rev. B}\ }\textbf {\bibinfo
  {volume} {102}},\ \bibinfo {pages} {045413} (\bibinfo {year}
  {2020})}\BibitemShut {NoStop}%
\bibitem [{\citenamefont {Liu}\ \emph {et~al.}(2019)\citenamefont {Liu},
  \citenamefont {Shabani},\ and\ \citenamefont {Mitra}}]{Liu2}%
  \BibitemOpen
  \bibfield  {author} {\bibinfo {author} {\bibfnamefont {D.~T.}\ \bibnamefont
  {Liu}}, \bibinfo {author} {\bibfnamefont {J.}~\bibnamefont {Shabani}},\ and\
  \bibinfo {author} {\bibfnamefont {A.}~\bibnamefont {Mitra}},\ }\href@noop {}
  {\bibfield  {journal} {\bibinfo  {journal} {Phys. Rev. B}\ }\textbf {\bibinfo
  {volume} {99}},\ \bibinfo {pages} {094303} (\bibinfo {year}
  {2019})}\BibitemShut {NoStop}%
\bibitem [{\citenamefont {Yang}\ \emph {et~al.}(2021)\citenamefont {Yang},
  \citenamefont {Yang}, \citenamefont {Hu},\ and\ \citenamefont {Liu}}]{Yang}%
  \BibitemOpen
  \bibfield  {author} {\bibinfo {author} {\bibfnamefont {Z.}~\bibnamefont
  {Yang}}, \bibinfo {author} {\bibfnamefont {Q.}~\bibnamefont {Yang}}, \bibinfo
  {author} {\bibfnamefont {J.}~\bibnamefont {Hu}},\ and\ \bibinfo {author}
  {\bibfnamefont {D.~E.}\ \bibnamefont {Liu}},\ }\href@noop {} {\bibfield
  {journal} {\bibinfo  {journal} {Phys. Rev. Lett.}\ }\textbf {\bibinfo
  {volume} {126}},\ \bibinfo {pages} {086801} (\bibinfo {year}
  {2021})}\BibitemShut {NoStop}%
\bibitem [{\citenamefont {Wingreen}\ \emph {et~al.}(1993)\citenamefont
  {Wingreen}, \citenamefont {Jauho},\ and\ \citenamefont {Meir}}]{Wingreen}%
  \BibitemOpen
  \bibfield  {author} {\bibinfo {author} {\bibfnamefont {N.~S.}\ \bibnamefont
  {Wingreen}}, \bibinfo {author} {\bibfnamefont {A.-P.}\ \bibnamefont
  {Jauho}},\ and\ \bibinfo {author} {\bibfnamefont {Y.}~\bibnamefont {Meir}},\
  }\href@noop {} {\bibfield  {journal} {\bibinfo  {journal} {Phys. Rev. B}\
  }\textbf {\bibinfo {volume} {48}},\ \bibinfo {pages} {8487} (\bibinfo {year}
  {1993})}\BibitemShut {NoStop}%
\bibitem [{\citenamefont {Sun}\ \emph {et~al.}(1999)\citenamefont {Sun},
  \citenamefont {Wang},\ and\ \citenamefont {Lin}}]{Sun1}%
  \BibitemOpen
  \bibfield  {author} {\bibinfo {author} {\bibfnamefont {Q.-f.}\ \bibnamefont
  {Sun}}, \bibinfo {author} {\bibfnamefont {J.}~\bibnamefont {Wang}},\ and\
  \bibinfo {author} {\bibfnamefont {T.-h.}\ \bibnamefont {Lin}},\ }\href@noop
  {} {\bibfield  {journal} {\bibinfo  {journal} {Phys. Rev. B}\ }\textbf
  {\bibinfo {volume} {59}},\ \bibinfo {pages} {13126} (\bibinfo {year}
  {1999})}\BibitemShut {NoStop}%
\bibitem [{\citenamefont {Sun}\ and\ \citenamefont {Lin}(1997)}]{Sun2}%
  \BibitemOpen
  \bibfield  {author} {\bibinfo {author} {\bibfnamefont {Q.-f.}\ \bibnamefont
  {Sun}}\ and\ \bibinfo {author} {\bibfnamefont {T.-h.}\ \bibnamefont {Lin}},\
  }\href@noop {} {\bibfield  {journal} {\bibinfo  {journal} {J. Phys.: Condens.
  Matter}\ }\textbf {\bibinfo {volume} {9}},\ \bibinfo {pages} {4875} (\bibinfo
  {year} {1997})}\BibitemShut {NoStop}%
\bibitem [{\citenamefont {Baran}\ and\ \citenamefont
  {Doma\ifmmode~\acute{n}\else \'{n}\fi{}ski}(2019)}]{Baran}%
  \BibitemOpen
  \bibfield  {author} {\bibinfo {author} {\bibfnamefont {B.}~\bibnamefont
  {Baran}}\ and\ \bibinfo {author} {\bibfnamefont {T.}~\bibnamefont
  {Doma\ifmmode~\acute{n}\else \'{n}\fi{}ski}},\ }\href@noop {} {\bibfield
  {journal} {\bibinfo  {journal} {Phys. Rev. B}\ }\textbf {\bibinfo {volume}
  {100}},\ \bibinfo {pages} {085414} (\bibinfo {year} {2019})}\BibitemShut
  {NoStop}%
\bibitem [{\citenamefont {Jauho}\ \emph {et~al.}(1994)\citenamefont {Jauho},
  \citenamefont {Wingreen},\ and\ \citenamefont {Meir}}]{Jauho}%
  \BibitemOpen
  \bibfield  {author} {\bibinfo {author} {\bibfnamefont {A.-P.}\ \bibnamefont
  {Jauho}}, \bibinfo {author} {\bibfnamefont {N.~S.}\ \bibnamefont
  {Wingreen}},\ and\ \bibinfo {author} {\bibfnamefont {Y.}~\bibnamefont
  {Meir}},\ }\href@noop {} {\bibfield  {journal} {\bibinfo  {journal} {Phys.
  Rev. B}\ }\textbf {\bibinfo {volume} {50}},\ \bibinfo {pages} {5528}
  (\bibinfo {year} {1994})}\BibitemShut {NoStop}%
\bibitem [{\citenamefont {Kells}\ \emph {et~al.}(2012)\citenamefont {Kells},
  \citenamefont {Meidan},\ and\ \citenamefont {Brouwer}}]{Kells}%
  \BibitemOpen
  \bibfield  {author} {\bibinfo {author} {\bibfnamefont {G.}~\bibnamefont
  {Kells}}, \bibinfo {author} {\bibfnamefont {D.}~\bibnamefont {Meidan}},\ and\
  \bibinfo {author} {\bibfnamefont {P.~W.}\ \bibnamefont {Brouwer}},\
  }\href@noop {} {\bibfield  {journal} {\bibinfo  {journal} {Phys. Rev. B}\
  }\textbf {\bibinfo {volume} {86}},\ \bibinfo {pages} {100503(R)} (\bibinfo
  {year} {2012})}\BibitemShut {NoStop}%
\bibitem [{\citenamefont {Vuik}\ \emph {et~al.}(2019)\citenamefont {Vuik},
  \citenamefont {Nijholt}, \citenamefont {Akhmerov},\ and\ \citenamefont
  {Wimmer}}]{Vuik}%
  \BibitemOpen
  \bibfield  {author} {\bibinfo {author} {\bibfnamefont {A.}~\bibnamefont
  {Vuik}}, \bibinfo {author} {\bibfnamefont {B.}~\bibnamefont {Nijholt}},
  \bibinfo {author} {\bibfnamefont {A.~R.}\ \bibnamefont {Akhmerov}},\ and\
  \bibinfo {author} {\bibfnamefont {M.}~\bibnamefont {Wimmer}},\ }\href@noop {}
  {\bibfield  {journal} {\bibinfo  {journal} {SciPost Phys.}\ }\textbf
  {\bibinfo {volume} {7}},\ \bibinfo {pages} {61} (\bibinfo {year}
  {2019})}\BibitemShut {NoStop}%
\bibitem [{\citenamefont {Prada}\ \emph {et~al.}(2020)\citenamefont {Prada},
  \citenamefont {San-Jose}, \citenamefont {de~Moor}, \citenamefont {Geresdi},
  \citenamefont {Lee}, \citenamefont {Klinovaja}, \citenamefont {Loss},
  \citenamefont {Nyg\r{a}rd}, \citenamefont {Aguado},\ and\ \citenamefont
  {Kouwenhoven}}]{Prada2}%
  \BibitemOpen
  \bibfield  {author} {\bibinfo {author} {\bibfnamefont {E.}~\bibnamefont
  {Prada}}, \bibinfo {author} {\bibfnamefont {P.}~\bibnamefont {San-Jose}},
  \bibinfo {author} {\bibfnamefont {M.~W.~A.}\ \bibnamefont {de~Moor}},
  \bibinfo {author} {\bibfnamefont {A.}~\bibnamefont {Geresdi}}, \bibinfo
  {author} {\bibfnamefont {E.~J.~H.}\ \bibnamefont {Lee}}, \bibinfo {author}
  {\bibfnamefont {J.}~\bibnamefont {Klinovaja}}, \bibinfo {author}
  {\bibfnamefont {D.}~\bibnamefont {Loss}}, \bibinfo {author} {\bibfnamefont
  {J.}~\bibnamefont {Nyg\r{a}rd}}, \bibinfo {author} {\bibfnamefont
  {R.}~\bibnamefont {Aguado}},\ and\ \bibinfo {author} {\bibfnamefont {L.~P.}\
  \bibnamefont {Kouwenhoven}},\ }\href@noop {} {\bibfield  {journal} {\bibinfo
  {journal} {Nat. Rev. Phys.}\ }\textbf {\bibinfo {volume} {2}},\ \bibinfo
  {pages} {575} (\bibinfo {year} {2020})}\BibitemShut {NoStop}%
\bibitem [{\citenamefont {\color{black}{C. O. Tabarner}}(2017)}]{Carlos}%
  \BibitemOpen
  \bibfield  {author} {\bibinfo {author} {\bibnamefont {\color{black}{C. O.
  Tabarner}}},\ }\href@noop {} {Master's thesis},\ \bibinfo  {school}
  {University of Copenhagen} (\bibinfo {year} {2017})\BibitemShut {NoStop}%
\bibitem [{\citenamefont {Rudner}\ and\ \citenamefont
  {Lindner}(2020)}]{Rudner}%
  \BibitemOpen
  \bibfield  {author} {\bibinfo {author} {\bibfnamefont {M.~S.}\ \bibnamefont
  {Rudner}}\ and\ \bibinfo {author} {\bibfnamefont {N.~H.}\ \bibnamefont
  {Lindner}},\ }\href@noop {} {\bibfield  {journal} {\bibinfo  {journal} {Nat.
  Rev. Phys}\ }\textbf {\bibinfo {volume} {2}},\ \bibinfo {pages} {229}
  (\bibinfo {year} {2020})}\BibitemShut {NoStop}%
\bibitem [{\citenamefont {Kohler}\ \emph {et~al.}(2005)\citenamefont {Kohler},
  \citenamefont {Lehmann},\ and\ \citenamefont {Hänggi}}]{Kohler}%
  \BibitemOpen
  \bibfield  {author} {\bibinfo {author} {\bibfnamefont {S.}~\bibnamefont
  {Kohler}}, \bibinfo {author} {\bibfnamefont {J.}~\bibnamefont {Lehmann}},\
  and\ \bibinfo {author} {\bibfnamefont {P.}~\bibnamefont {Hänggi}},\
  }\href@noop {} {\bibfield  {journal} {\bibinfo  {journal} {Phys. Rep.}\
  }\textbf {\bibinfo {volume} {406}},\ \bibinfo {pages} {379} (\bibinfo {year}
  {2005})}\BibitemShut {NoStop}%
\bibitem [{\citenamefont {Li}\ \emph {et~al.}(2018)\citenamefont {Li},
  \citenamefont {Song}, \citenamefont {Liu}, \citenamefont {Jiang},
  \citenamefont {Sun},\ and\ \citenamefont {Xie}}]{Li}%
  \BibitemOpen
  \bibfield  {author} {\bibinfo {author} {\bibfnamefont {Y.-H.}\ \bibnamefont
  {Li}}, \bibinfo {author} {\bibfnamefont {J.}~\bibnamefont {Song}}, \bibinfo
  {author} {\bibfnamefont {J.}~\bibnamefont {Liu}}, \bibinfo {author}
  {\bibfnamefont {H.}~\bibnamefont {Jiang}}, \bibinfo {author} {\bibfnamefont
  {Q.-F.}\ \bibnamefont {Sun}},\ and\ \bibinfo {author} {\bibfnamefont {X.~C.}\
  \bibnamefont {Xie}},\ }\href@noop {} {\bibfield  {journal} {\bibinfo
  {journal} {Phys. Rev. B}\ }\textbf {\bibinfo {volume} {97}},\ \bibinfo
  {pages} {045423} (\bibinfo {year} {2018})}\BibitemShut {NoStop}%
\bibitem [{\citenamefont {Cuevas}\ \emph {et~al.}(1996)\citenamefont {Cuevas},
  \citenamefont {Mart\'{\i}n-Rodero},\ and\ \citenamefont {Yeyati}}]{Cuevas}%
  \BibitemOpen
  \bibfield  {author} {\bibinfo {author} {\bibfnamefont {J.~C.}\ \bibnamefont
  {Cuevas}}, \bibinfo {author} {\bibfnamefont {A.}~\bibnamefont
  {Mart\'{\i}n-Rodero}},\ and\ \bibinfo {author} {\bibfnamefont {A.~L.}\
  \bibnamefont {Yeyati}},\ }\href@noop {} {\bibfield  {journal} {\bibinfo
  {journal} {Phys. Rev. B}\ }\textbf {\bibinfo {volume} {54}},\ \bibinfo
  {pages} {7366} (\bibinfo {year} {1996})}\BibitemShut {NoStop}%
\bibitem [{\citenamefont {Das~Sarma}\ \emph {et~al.}(2012)\citenamefont
  {Das~Sarma}, \citenamefont {Sau},\ and\ \citenamefont {Stanescu}}]{Das2}%
  \BibitemOpen
  \bibfield  {author} {\bibinfo {author} {\bibfnamefont {S.}~\bibnamefont
  {Das~Sarma}}, \bibinfo {author} {\bibfnamefont {J.~D.}\ \bibnamefont {Sau}},\
  and\ \bibinfo {author} {\bibfnamefont {T.~D.}\ \bibnamefont {Stanescu}},\
  }\href@noop {} {\bibfield  {journal} {\bibinfo  {journal} {Phys. Rev. B}\
  }\textbf {\bibinfo {volume} {86}},\ \bibinfo {pages} {220506(R)} (\bibinfo
  {year} {2012})}\BibitemShut {NoStop}%
\bibitem [{Note1()}]{Note1}%
  \BibitemOpen
  \bibinfo {note} {The fitting for square ratio of Bessel function may be
  multivalued. But we can avoid this ambiguity by considering a small field
  strength and $Q<1$}\BibitemShut {NoStop}%
\bibitem [{\citenamefont {Serina}\ \emph {et~al.}(2018)\citenamefont {Serina},
  \citenamefont {Loss},\ and\ \citenamefont {Klinovaja}}]{Serina}%
  \BibitemOpen
  \bibfield  {author} {\bibinfo {author} {\bibfnamefont {M.}~\bibnamefont
  {Serina}}, \bibinfo {author} {\bibfnamefont {D.}~\bibnamefont {Loss}},\ and\
  \bibinfo {author} {\bibfnamefont {J.}~\bibnamefont {Klinovaja}},\ }\href@noop
  {} {\bibfield  {journal} {\bibinfo  {journal} {Phys. Rev. B}\ }\textbf
  {\bibinfo {volume} {98}},\ \bibinfo {pages} {035419} (\bibinfo {year}
  {2018})}\BibitemShut {NoStop}%
\bibitem [{\citenamefont {Wiedenmann}\ \emph {et~al.}(2016)\citenamefont
  {Wiedenmann}, \citenamefont {Bocquillon}, \citenamefont {Deacon},
  \citenamefont {Hartinger}, \citenamefont {Herrmann}, \citenamefont
  {Klapwijk}, \citenamefont {Maier}, \citenamefont {Ames}, \citenamefont
  {Brüne}, \citenamefont {Gould}, \citenamefont {Oiwa}, \citenamefont
  {Ishibashi}, \citenamefont {Tarucha}, \citenamefont {Buhmann},\ and\
  \citenamefont {Molenkamp}}]{Wiedenmann}%
  \BibitemOpen
  \bibfield  {author} {\bibinfo {author} {\bibfnamefont {J.}~\bibnamefont
  {Wiedenmann}}, \bibinfo {author} {\bibfnamefont {E.}~\bibnamefont
  {Bocquillon}}, \bibinfo {author} {\bibfnamefont {R.~S.}\ \bibnamefont
  {Deacon}}, \bibinfo {author} {\bibfnamefont {S.}~\bibnamefont {Hartinger}},
  \bibinfo {author} {\bibfnamefont {O.}~\bibnamefont {Herrmann}}, \bibinfo
  {author} {\bibfnamefont {T.~M.}\ \bibnamefont {Klapwijk}}, \bibinfo {author}
  {\bibfnamefont {L.}~\bibnamefont {Maier}}, \bibinfo {author} {\bibfnamefont
  {C.}~\bibnamefont {Ames}}, \bibinfo {author} {\bibfnamefont {C.}~\bibnamefont
  {Brüne}}, \bibinfo {author} {\bibfnamefont {C.}~\bibnamefont {Gould}},
  \bibinfo {author} {\bibfnamefont {A.}~\bibnamefont {Oiwa}}, \bibinfo {author}
  {\bibfnamefont {K.}~\bibnamefont {Ishibashi}}, \bibinfo {author}
  {\bibfnamefont {S.}~\bibnamefont {Tarucha}}, \bibinfo {author} {\bibfnamefont
  {H.}~\bibnamefont {Buhmann}},\ and\ \bibinfo {author} {\bibfnamefont {L.~W.}\
  \bibnamefont {Molenkamp}},\ }\href@noop {} {\bibfield  {journal} {\bibinfo
  {journal} {Nat. Commun.}\ }\textbf {\bibinfo {volume} {7}},\ \bibinfo {pages}
  {10303} (\bibinfo {year} {2016})}\BibitemShut {NoStop}%
\bibitem [{\citenamefont {Peters}\ \emph {et~al.}(2020)\citenamefont {Peters},
  \citenamefont {Bogdanoff}, \citenamefont {Acero~González}, \citenamefont
  {Melischek}, \citenamefont {Simon}, \citenamefont {Reecht}, \citenamefont
  {Winkelmann}, \citenamefont {von Oppen},\ and\ \citenamefont
  {Franke}}]{Peters}%
  \BibitemOpen
  \bibfield  {author} {\bibinfo {author} {\bibfnamefont {O.}~\bibnamefont
  {Peters}}, \bibinfo {author} {\bibfnamefont {N.}~\bibnamefont {Bogdanoff}},
  \bibinfo {author} {\bibfnamefont {S.}~\bibnamefont {Acero~González}},
  \bibinfo {author} {\bibfnamefont {L.}~\bibnamefont {Melischek}}, \bibinfo
  {author} {\bibfnamefont {J.~R.}\ \bibnamefont {Simon}}, \bibinfo {author}
  {\bibfnamefont {G.}~\bibnamefont {Reecht}}, \bibinfo {author} {\bibfnamefont
  {C.~B.}\ \bibnamefont {Winkelmann}}, \bibinfo {author} {\bibfnamefont
  {F.}~\bibnamefont {von Oppen}},\ and\ \bibinfo {author} {\bibfnamefont
  {K.~J.}\ \bibnamefont {Franke}},\ }\href@noop {} {\bibfield  {journal}
  {\bibinfo  {journal} {Nat. Phys.}\ }\textbf {\bibinfo {volume} {16}},\
  \bibinfo {pages} {1222} (\bibinfo {year} {2020})}\BibitemShut {NoStop}%
\bibitem [{\citenamefont {van Zanten}\ \emph {et~al.}(2020)\citenamefont {van
  Zanten}, \citenamefont {Sabonis}, \citenamefont {Suter}, \citenamefont
  {V$\ddot{a}$yrynen}, \citenamefont {Karzig}, \citenamefont {Pikulin},
  \citenamefont {O'Farrell}, \citenamefont {Razmadze}, \citenamefont
  {Petersson}, \citenamefont {Krogstrup},\ and\ \citenamefont
  {Marcus}}]{Zanten}%
  \BibitemOpen
  \bibfield  {author} {\bibinfo {author} {\bibfnamefont {D.~M.~T.}\
  \bibnamefont {van Zanten}}, \bibinfo {author} {\bibfnamefont
  {D.}~\bibnamefont {Sabonis}}, \bibinfo {author} {\bibfnamefont
  {J.}~\bibnamefont {Suter}}, \bibinfo {author} {\bibfnamefont {J.~I.}\
  \bibnamefont {V$\ddot{a}$yrynen}}, \bibinfo {author} {\bibfnamefont
  {T.}~\bibnamefont {Karzig}}, \bibinfo {author} {\bibfnamefont {D.~I.}\
  \bibnamefont {Pikulin}}, \bibinfo {author} {\bibfnamefont {E.~C.~T.}\
  \bibnamefont {O'Farrell}}, \bibinfo {author} {\bibfnamefont {D.}~\bibnamefont
  {Razmadze}}, \bibinfo {author} {\bibfnamefont {K.~D.}\ \bibnamefont
  {Petersson}}, \bibinfo {author} {\bibfnamefont {P.}~\bibnamefont
  {Krogstrup}},\ and\ \bibinfo {author} {\bibfnamefont {C.~M.}\ \bibnamefont
  {Marcus}},\ }\href@noop {} {\bibfield  {journal} {\bibinfo  {journal} {Nat.
  Phys.}\ }\textbf {\bibinfo {volume} {16}},\ \bibinfo {pages} {663} (\bibinfo
  {year} {2020})}\BibitemShut {NoStop}%
\bibitem [{\citenamefont {Pan}\ and\ \citenamefont {Das~Sarma}(2020)}]{Pan}%
  \BibitemOpen
  \bibfield  {author} {\bibinfo {author} {\bibfnamefont {H.}~\bibnamefont
  {Pan}}\ and\ \bibinfo {author} {\bibfnamefont {S.}~\bibnamefont
  {Das~Sarma}},\ }\href@noop {} {\bibfield  {journal} {\bibinfo  {journal}
  {Phys. Rev. Research}\ }\textbf {\bibinfo {volume} {2}},\ \bibinfo {pages}
  {013377} (\bibinfo {year} {2020})}\BibitemShut {NoStop}%
\bibitem [{\citenamefont {Moore}\ \emph {et~al.}(2018)\citenamefont {Moore},
  \citenamefont {Stanescu},\ and\ \citenamefont {Tewari}}]{Moore2}%
  \BibitemOpen
  \bibfield  {author} {\bibinfo {author} {\bibfnamefont {C.}~\bibnamefont
  {Moore}}, \bibinfo {author} {\bibfnamefont {T.~D.}\ \bibnamefont
  {Stanescu}},\ and\ \bibinfo {author} {\bibfnamefont {S.}~\bibnamefont
  {Tewari}},\ }\href@noop {} {\bibfield  {journal} {\bibinfo  {journal} {Phys.
  Rev. B}\ }\textbf {\bibinfo {volume} {97}},\ \bibinfo {pages} {165302}
  (\bibinfo {year} {2018})}\BibitemShut {NoStop}%
\bibitem [{\citenamefont {Stanescu}\ and\ \citenamefont
  {Tewari}(2019)}]{Stanescu}%
  \BibitemOpen
  \bibfield  {author} {\bibinfo {author} {\bibfnamefont {T.~D.}\ \bibnamefont
  {Stanescu}}\ and\ \bibinfo {author} {\bibfnamefont {S.}~\bibnamefont
  {Tewari}},\ }\href@noop {} {\bibfield  {journal} {\bibinfo  {journal} {Phys.
  Rev. B}\ }\textbf {\bibinfo {volume} {100}},\ \bibinfo {pages} {155429}
  (\bibinfo {year} {2019})}\BibitemShut {NoStop}%
\end{thebibliography}%
\end{document}